\documentclass[pra,notitlepage]{revtex4-1}
\usepackage{graphicx}
\usepackage{sidecap}
\usepackage{caption,subcaption}
\usepackage{xcolor}
\usepackage{amsmath}
\usepackage{amsfonts}
\usepackage{titlesec}
\usepackage{tabu}
\usepackage{caption}
\usepackage{amsthm,amssymb,cases}
\usepackage{bm}
\usepackage{here}

\newcommand{\ket}[1]{|#1\rangle}

\renewcommand{\Re}{{\rm Re}\,}
\renewcommand{\Im}{{\rm Im}\,}
\newcommand{\tr}{{\rm tr}\,}

\begin{document}
\title{Robust phase estimation of Gaussian states in the presence of outlier quantum states}
\author{Yukito Mototake}\email{mototake@mail.uec.jp}
\author{Jun Suzuki}\email{junsuzuki@uec.ac.jp}
\date{\today}

\affiliation{
Graduate School of Informatics and Engineering, The University of Electro-Communications,\\
1-5-1 Chofugaoka, Chofu-shi, Tokyo, 182-8585 Japan
}
\begin{abstract}
In this paper, we investigate the problem of estimating the phase of a coherent state in the presence of unavoidable noisy quantum states. These unwarranted quantum states are represented by outlier quantum states in this study. 
We first present a statistical framework of robust statistics in a quantum system to handle outlier quantum states.
We then apply the method of M-estimators to suppress untrusted measurement outcomes due to outlier quantum states. 
Our proposal has the advantage over the classical methods in being systematic, easy to implement, and robust against occurrence of noisy states.
\end{abstract}

% Keywords
%\keyword{robust statistics; phase estimation; quantum gaussian state; outlier quantum state}
%%%%%%%%%%%%%%%%%%%%%%%%%%%%%%%%%%%%%%%%%%
%%%%%%%%%%%%%%%%%%%%%%%%%%%%%%%%%%%%%%%%%%
\maketitle

\section{Introduction}
One of the challenges in developing quantum information technologies is to suppress uncontrollable elements both in classical and quantum devices.
This has been an active research subject under the name of state preparation and measurement (SPAM) error \cite{merkel2013self,ferrie2014self,sugiyama2018reliable}.
For example, imperfection at the quantum state preparation stage generates unwarranted outlier quantum states at random.
Under this circumstance, it is impossible to predict precisely when these outlier quantum states are generated.
The resulting quantum state is represented by a convex mixture of the desired quantum state and the outlier quantum states. Thus, the actual model is contaminated by outlier quantum states. 
No matter how small the occurrence of these outlier quantum states is, they affect statistics of measurement outcomes.
In this paper, we develop a statistical framework to handle certain types of SPAM errors by applying robust statistics \cite{huber2004robust,andrews2015robust,wilcox2011introduction,hampel2011robust,maronna2019robust}.

Data due to undesired quantum states are called {\it outliers} in statistics, and they typically lie on outside the range of trusted data.
The traditional working rule to remove outliers in practice is so-called the 3-$\sigma$ rule.
In short, we discard all the data, which are 3-$\sigma$ away from the sample mean.
This is intimately related to the tradition of 3-$\sigma$ confidence in statistics.
However, there is no justification for such heuristic and subjective data processing from the statistical point of view.
The current status in the community is in fact that we should not rely on the 3-$\sigma$ confidence \cite{wasserstein2016pvalue,camerer2018evaluating,wasserstein2019beyond}.
Robust statistics is a branch of statistics and has been one of the proper tools to remove outliers systematically \cite{huber2004robust,andrews2015robust,wilcox2011introduction,hampel2011robust,maronna2019robust}.
Recent success of robust statistics in machine learning is another justification to use it rather than the 3-$\sigma$ rule.

The problem of handling outlier quantum states is a practical and important issue in any quantum communication protocols.
Yet, it seems that this problem has not been addressed properly in the framework of robust statistics to our best knowledge.
The main contribution of this paper is first to present a statistical framework to handle the problem of state estimation in the presence of unknown outlier quantum states.
We then demonstrate usefulness and effectiveness of robust statistics in the quantum case.
To this end, we consider a specific problem; phase estimation of coherent states in the presence of outlier quantum states.
This problem has many practical applications in quantum communication protocols using continuous variables \cite{braunstein2005quantum,wang2007quantum,serafini2017quantum}.
Measurement data drawn according to this contaminated state contain outliers.
We apply two specific M-estimators to make robust estimates from data.
We find that a recently proposed robust estimator in Ref.~\cite{fujisawa2008robust} based on the generalized divergence performs well.
We compare its performance to other methods and evaluate its robustness based on our proposed figure of merit called $\varepsilon$-curve.

The system of noisy quantum gaussian states has a long history
\cite{helstrom1968minimum,helstrom1969quantum,helstrom1974noncommuting,yuen1973multiple,helstrom,holevo}.
Further developments in applications to quantum information processing protocols were studied over the last two decades
\cite{fujiwara1999estimation,d2000parameter,braunstein2005quantum,wang2007quantum,serafini2017quantum,helstrom,holevo,QSEbook,hayashi2017quantum}
We note that our model is described by a different class of noise models studied in literature.
The main difference from the previous studies is that our noise model is not a typical completely positive and trace preserving (CP-TP) map.
In particular, the model studied in this paper is not described by a unital CP-TP map, but a mixture of different quantum gaussian states.
Another distinction is that we do not need to assume a specific form for outlier quantum states when applying our method to real data. 
Note that we could apply the conventional parameter estimation method, if we know specific forms of noisy quantum states. For example, we can treat noisy states as nuisance \cite{suzuki2020nuisance,suzuki2020quantum}. 
In contrast, the proposed method of M-estimators can be applied to the occurrence of unknown noisy quantum states. This is one of the practical advantages of the theory of robust statistics. In this sense, we do not need to know any physical mechanism to create these outlier quantum states in our setting.

In our study, the question of the ultimate precision limit, which is an active subject of quantum metrology in quantum gaussian systems
\cite{aspachs2009phase,pinel2013quantum,bradshaw2018ultimate,oh2019optimal,lee2019using,Arnhem_2019,oh2020optimal,assad2020accessible}, is not a primary objective.
This is because the ultimate limit typically depends on the nature of the noise model.
Furthermore, we cannot derive such the ultimate precision limit without having knowledge on outlier quantum states.
Instead, we are aiming at finding a practical and robust estimation strategy, and this is the basic philosophy of robust statistics.

The outline of the paper is as follows.
In Section~\ref{sec2}, a short summary of robust statistics is given for the paper begin self-contained.
Section \ref{sec:rubustness} discusses robustness of estimators and propose a new measure of robustness used in this paper.
In Section~\ref{sec4}, we develop the concept of a quantum statistical model in the presence of outlier quantum states.
In Section~\ref{sec5}, we apply our formalism to the problem of phase estimation of coherent states.
The last Section~\ref{sec6} gives a summary of the paper.

\section{Preliminaries}\label{sec2}
In this section, we give a short summary on robust statistics base on the theory of M-estimator. To provide the basic idea, we focus on estimating a single parameter case. Its extension to multiple parameters can be done similarly. 
The purpose of this section is to give a simple recipe to apply M-estimators. Readers who are only interested in applying M-estimators can skip most of this section. We provide a summary of how to apply the theory of M-estimators at the end of this section. 
See Refs. \cite{huber2004robust,andrews2015robust,wilcox2011introduction,hampel2011robust,maronna2019robust} for more detailed discussions.

\subsection{M-estimator}\label{sec:m-est}
An M-estimator is a generalization of the maximum likelihood estimator (MLE) and is defined as follows.
Consider a datum $\mathcal{X}=\left\{x_{1}, \ldots, x_{n}\right\}$ of the sample size $n$,
which is identically and independently distributed (i.i.d.) according to one-parameter family of probability density functions $ f(x | \theta)$.
The MLE to estimate $\theta$ is defined by
\begin{equation}
\hat{\theta}_{\rm MLE}=\arg \max _{\theta} \sum_{i=1}^{n} \log f \left(x_{i} | \theta\right).
\end{equation}
The MLE needs to be a stationary point of the logarithmic likelihood equation:
\begin{equation}
\sum_{i=1}^{n} \frac{d}{d \theta} \log f(x_{i} | \theta) = 0.
\end{equation}
The basic philosophy behind the M-estimator is to generalize this equation by
\begin{equation} \label{psiEq}
\sum_{i=1}^{n} \psi\left(x_{i} | \theta\right)=0,
\end{equation}
where $\psi(x | \theta)$ can be an arbitrary function as long as it satisfies a certain conditions \cite{maronna2019robust}.
An estimator, which is defined the above generalized stationary condition \eqref{psiEq}, is called an {\it M-estimator}.
Equation \eqref{psiEq} is called an {\it M-equation} in robust statistics.
Clearly, M-estimator depends on a choice of $\psi$ function.
The choice $\psi(x | \theta)=\frac{d}{d \theta} \log f(x | \theta)$ corresponds to the familiar MLE.

In the following discussion, we consider estimation of the expectation value $\mu$ of the model for simplicity.
We assume that the true probability density function is a function of $f(x-\mu)$ and is symmetric at the origin,
i.e., $f(-x)=f(x)$.  A typical example of this kind is the normal distribution.
Note that it is easy to generalize our setting to an arbitrary location model \cite{maronna2019robust}.
Under this assumption, an M-equation to infer the parameter $\mu$ is a function of $x-\mu$, and hence we have
\begin{equation*}
\sum_{i=1}^{n} \psi\left(x_{i}-\mu\right)=0.
\end{equation*}
There is the zoology of M-estimators studied in robust statistics, see for example Refs.~\cite{andrews2015robust,hampel2011robust,maronna2019robust}.
For our purpose, we consider two specific M-estimators; bisquare and gamma M-estimators.
The $\psi$ function for bisquare M-estimator (also known as Tukey's estimator) is given by
\begin{equation}
\psi_{\rm bi}(x-\mu)=\begin{cases}
{(x-\mu)\left\{1-\left(\frac{x-\mu}{c}\right)^{2}\right\}^{2}} &\quad {(|x-\mu| \leq c)} \\[2ex]
{0} &\quad {(|x-\mu|>c)}\end{cases},
\end{equation}
where $c$ is a tuning parameter. The basic property of bisquare M-estimator is to suppress
contribution from data which are far away from the true parameter $\mu$. 
Clearly, $\psi_{\rm bi}$ function vanishes at $|x-\mu|=c$ smoothly.

In Ref.~\cite{fujisawa2008robust}, an M-estimator was proposed based on the gamma divergence, which is a generalization of the Kullback-Leibler divergence (also called the relative entropy).
Its performance was demonstrated to be more robust than the traditional M-estimators.
When the true probability density function obeys the normal distribution, the $\psi$ function is defined by
\begin{equation}
  \psi_{\rm gam}(x-\mu) =  \left[\frac{1}{\sqrt{2 \pi \sigma^{2}}} \exp \left\{-\frac{1}{2 \sigma^{2}}(x-\mu)^{2}\right\}\right]^{\gamma}(x-\mu),
\end{equation}
where $\gamma>0$ is a tuning parameter appearing in the power of the normal distribution ${\cal N}(\mu,\sigma)$ ($\mu$ is the expectation value, and $\sigma$ is the standard deviation.).
Another tuning parameter, the standard deviation $\sigma$, needs to be properly chosen as well.
This will be discussed later. 
$\psi_{\rm gam}$ approaches $0$ as $|x|\to\infty$, and its convergence is exponential. 

\subsection{Tuning parameter}
We now discuss the issue of the tuning parameter.
Generally speaking, it is favorable to have less freedom for tuning parameters.
The performance of an M-estimator should also be independent on a choice of tuning parameters.
However, there is no universally accepted methodology to setup these tuning parameters. 
One of the standards is based on asymptotic relative efficiency as follows. 
This guarantees the M-estimator would perform well in the large sample regime. 
Let us assume that the true model is the normal distribution, and consider an M-estimator.
Asymptotic relative efficiency $\eta$ is defined by the ratio,
\begin{equation}
\eta = \frac{\sigma_{\rm MLE}^2}{\sigma_{\rm asymptotic}^2},
\end{equation}
where $\sigma^2_{\rm MLE}$ and $\sigma^2_{\rm asymptotic}$ denote asymptotic variances of the MLE and the M-estimator, respectively.
Following the tradition, tuning parameters are determined by the condition $\eta=0.95$ \cite{maronna2019robust}.
For bisquare and gamma M-estimators, explicit values for the tuning parameters are known as 
\begin{align*}
c&=4.68\, \hat{\sigma} \quad\mbox{bisquare},\\
\gamma&=0.2 \quad\mbox{gamma},
\end{align*}
where $\hat{\sigma}$ is an estimated standard deviation and is often set to the normalized version of the median absolute deviation, called the MADN. 
(The MADN is defined by MAD/0.675 with MAD the median absolute deviation.)
Another tuning parameter of gamma M-estimator is set to be the estimate of the standard deviation.
A remark concerning gamma M-estimator is important.
When outliers are not far away from neighborhood of true data, the choice $\gamma=0.5$ is observed to be best from many examples known in literature.
We also analyzed this peculiar trick for several gaussian models and reached the same conclusion.
Therefore, we adopt the choice $\gamma=0.5$ in the rest of the paper.

\subsection{Iterative algorithm}\label{sec:IterAlg}
The M-equation \eqref{psiEq} is a non-linear function in general, and there is no efficient way to find root of the equation.
This point will be more problematic when estimating multiple-parameters, since one has to solve a coupled multivariate equations.
A simple iterative method is usually used in robust statistics to find an approximated solution \cite{maronna2019robust}.
Let us rewrite Eq.~\eqref{psiEq} as
\begin{align*}
0&=\sum_{i=1}^{n} \psi (x_{i}-\mu)\\
&=\sum_{i=1}^{n} W(x_{i}-\mu)\,\left(x_{i}-\mu\right),
\end{align*}
where $W(x)=\frac{\psi(x)}{x}$.
This can be put into the form,
\begin{equation}
\mu=\frac{\sum_{i=1}^{n} W\left(x_{i}-\mu\right) x_{i}}{\sum_{i=1}^{n} W\left(x_{i}-\mu\right)}.
\end{equation}
We start with an initial choice for $\mu^{(0)}$ and then iterate it according to
\begin{equation}
\mu^{(a+1)}=\frac{\sum_{i=1}^{n} W\left(x_{i}-\mu^{(a)}\right) x_{i}}{\sum_{i=1}^{n} W\left(x_{i}-\mu^{(a)}\right)}.
\end{equation}
After several iteration steps ($a=1,2,\ldots,a_{\rm fin}$), we stop the algorithm. 
Alternatively, any stopping rule can be adopted. 
A common choice for the initial value is the median when estimating the expectation value. 
As explicitly demonstrated in Section~\ref{sec:num_robust}, this iteration algorithm is efficient. 

We summarize the method of M-estimators in practice. 
First, we choose an appropriate $\psi$ function to apply. 
Second, we setup the M-equation by plugging an observed datum. 
Third, we solve this M-equation for the parameters of interest. 
One of them is to use the above simple iterative algorithm, but other methods can be applied as well. 
The obtained value is the estimate based on this M-estimator. 
As the last remark, it is better to apply several M-estimators and compare them. 
Based on comparison, we adjust tuning parameters of the chosen $\psi$ function. 
Repeating this procedure, we can obtain a more reliable estimate. 

\section{Robustness of M-estimator} \label{sec:rubustness}
In robust statistics, one of the main objectives is to construct an estimator, which is not affected by outliers.
This then lead the study of robustness of M-estimators.
There are several known measures for evaluating robustness quantitatively such as 
influence curve, gross error sensitivity, local shift sensitivity, break down point and so on \cite{maronna2019robust}.
In the following, we first explain the most common figure of merit for robustness, a breakdown point,
and then we introduce a new quantity $\varepsilon$-curve for our purpose. 
Readers who are not interested in detail can skip Section 3.2 and 3.3. 

\subsection{Classical contaminated model}
We now describe a classical statistical model in the presence of outliers, 
which is known as the contaminated model in robust statistics. 
Suppose we are interested in estimating $d$-parameter family of probability distributions $f_\theta$:
\begin{equation}\label{ideal_cmodel}
M=\{f_\theta| \theta = (\theta_1,\theta_2,\ldots,\theta_d)\in\Theta\},
\end{equation}
where $\Theta\subset \mathbb{R}^d$ is an open subset.
We denote its cumulative distribution function (CDF) by $F_\theta$.
The simplest situation is when possible outliers are generated by a single probability distribution $g$ whose CDF is $G$.
A probabilistic mixture of two probability distributions $f_\theta$ and $g$ gives the contaminated model:
\begin{equation}
f_\theta^\varepsilon=(1-\varepsilon) f_\theta+\varepsilon g,
\end{equation}
where $\varepsilon\in [0,1)$ denotes the strength of occurrence of outliers.
The strength of the noise $\varepsilon$ is usually referred to as the {\it contamination parameter}.
A familiar and classical example of this kind is to consider the normal distribution
$f_\theta=f^{\cal N}_{\mu,\sigma}$ with the expectation value $\mu$ and the standard deviation $\sigma$
as the ideal probability density function. Outliers are generated by another normal distribution whose expectation value is order of magnitude different from $\mu$. 
This is equivalent to a statistical model of the form:
\begin{equation}
M_{\rm Contaminated} =\{f^\varepsilon_\theta| \theta\in\Theta,\varepsilon\in[0,1)\}.
\end{equation}
Note that the contamination parameter as well as parameters characterizing in $g$ are to be regarded as nuisance parameters of the model $M_{\rm Contaminated}$.

The more general case of modeling outliers in the ideal case \eqref{ideal_cmodel} is introduced as follows.
Let ${\cal G}_{\rm outlier}$ be a set of all possible CDFs for generating outliers.
Its each element $G\in {\cal G}_{\rm outlier}$ corresponds to a different CDF generating different types of outliers.
The general contaminated model with outliers is described by the CDF,
\begin{equation}\label{contam_model}
F_\theta^\varepsilon=(1-\varepsilon) F_\theta+\varepsilon G\quad (G\in {\cal G}_{\rm outlier}),
\end{equation}
where $\varepsilon$ is the contamination parameter. 
In this general case, we do not need to assume specific forms of $G$ in order to apply the method of M-estimators. This is in contrast to the setting of parametric models. 

\subsection{Asymptotic breakdown point}
Intuitively, the value of the breakdown point represents the maximum ratio of outliers in a datum,
which can be suppressed by an M-estimator. 
Beyond this breakdown point, the M-estimator will no longer return a trusted value, which is typically infinitely large. 
Consider a one-parameter model $M=\{f_\theta| \theta\in\Theta\}$ for the sake of simplicity and
denote $ \partial \Theta$ by the boundary of the open set $\Theta$.
Fix an M-estimator $\hat{\theta}$ under consideration.
A contaminated model with outliers is given by \eqref{contam_model}. 
%\begin{equation}%\label{contam_model}
%F_\theta^\varepsilon=(1-\varepsilon) F_\theta+\varepsilon G,
%\end{equation}
%where $\varepsilon$ is the ratio of outliers and $G$ is a distribution generating outliers.

The asymptotic breakdown point of the estimator $\hat{\theta}$ for the model \eqref{contam_model} is
denoted by $\varepsilon^{*}\left(\hat {\theta},F_\theta\right)$.
This is defined by the maximum value of $\varepsilon\in [0,1)$ in which $\hat{\theta}$ on the boundary remains finite for arbitrary outlier distributions. Mathematically, this is expressed as
\begin{equation}
\forall \varepsilon < \varepsilon ^ { * },\forall G\in{\cal G}_{\rm outlier},\ \hat { \theta } _ { \infty } \left( ( 1 - \varepsilon ) F_\theta + \varepsilon G \right) \in K,
\end{equation}
where $K \subset \Theta$ is a closed bounded subset satisfying $K \cap \partial \Theta = \emptyset$ and
$\hat { \theta } _ { \infty }(F)$ denotes asymptotic behavior of $\hat { \theta }$ for the distribution $F$.
For example, the asymptotic breakdown point of the median is $0.5$,
since it gives values outside of the parameter set $\Theta$ when the half of the sample size are outliers.

\subsection{Finite breakdown point}
The above asymptotic breakdown point is defined by the asymptotic behavior of the M-estimator.
To evaluate this quantity for finite sample size data, there are several variants known in literature.
For our study, we focus on the finite breakdown point (FBP) based on replacement of the true data \cite{donoho1983notion,huber1984finite}.

Let $\hat {\theta} _ {n}$ be an M-estimator and consider a datum $\mathbf{x}=\left\{x_{1},x_2 \ldots, x_{n}\right\}$ of the sample size $n$.
Denote by $\mathbf{x};\mathbf{y}_m$ one of possible data obtained by replacing $n-m$ elements of $\mathbf{x}$ by $\mathbf{y}_m=\{y_1,y_2,\ldots,y_m\}$.
Intuitively, the FBP of the estimator $\hat {\theta} _ {n}$ for $\mathbf{x}$ is defined by
the maximum ratio $\frac{m}{n}$, in which $\hat {\theta} _ {n}(\mathbf{x};\mathbf{y}_m)$ behaves normal.
This FBP is denoted by $\varepsilon_{n}^{*}\left(\hat{\theta}_{n}, \mathbf{x}\right)$.
Typically, $\varepsilon_{n} ^ {*}(\hat{\theta}_{n}, \mathbf{x})$ is independent of the datum $ \mathbf {x}$.
It can be proven that the FBP converges to the asymptotic BP in the limit $ n \rightarrow \infty$ \cite{maronna2019robust}.

Formal definition of the FBP is as follows.
Let $ \mathcal{X}^{m}_{\mathbf{x}}$ be a set of all possible data whose intersection with $\mathbf{x}$ is $n-m$, i.e.,
\begin{equation*}
\mathcal{X}^{m}_{\mathbf{x}}=\{\mathbf{y}: \#(\mathbf{y})=n, \#(\mathbf{x} \cap \mathbf{y})=n-m\}.
\end{equation*}
The FBP for $\mathbf{x}$ is given by
\begin{align} \label{def_FBP}
&\varepsilon_{n}^{*}\left(\hat{\theta}_{n}, \mathbf{x}\right)
=\frac{m*}{n},\\
&m^*= \max \left\{m \geq 0 : \forall \mathbf{y} \in \mathcal{X}^{m}_{\mathbf{x}},\,
\hat{\theta}_{n}(\mathbf{y}) \notin \partial \Theta\cup \{\pm \infty \}\right\}. \nonumber
\end{align}
In our simulation for the FBP, we randomly generate outliers obeying the normal distribution whose expectation value has a different order from that of the true distribution. 
See Section~\ref{sec:num_robust} for details. 
By definition, the concept of the FBP relies on replacement of the actual data by artificially created outliers. 

Before closing this section, we discuss briefly the evaluation of robustness. 
The FBP seems to be the most common choice of robustness in literature. 
In statistics, one compares FBPs to show its robustness, when one proposes a new M-estimator. 
However, there were severe critiques to rely on it as a measure of robustness \cite{huber2004robust}.
One of the major objections is that this quantity does not concern the actual estimate at all. 
In other words, a robust estimator in the sense of high FBP could be a very poor estimator. 
To see this point, we will evaluate robustness based on the FBP together with a newly proposed figure of merit, called an $\varepsilon$-curve. 
This is defied by the behavior of M-estimators as a function of the contamination parameter $\varepsilon$. 
This $\varepsilon$-curve is conceptually simple, and it concerns the actual estimate. 
In this paper, we find it more natural to measure robustness based on the $\varepsilon$-curve. 
See Section~\ref{sec:num_robust} for a detailed comparison. 

\section{Quantum statistical model with outliers}\label{sec4}
In this section, we develop the concept of a quantum gaussian model with unavoidable outlier states. 
First, we consider the general quantum statistical model in the presence of outlier quantum states. 

\subsection{Quantum statistical model with outlier quantum states}
We assume that the true quantum state is characterized by a $d$-parameter $\theta=(\theta_1,\theta_2,\ldots,\theta_d)\in\Theta$.
The ideal quantum statistical model is given by the family of states
\begin{equation}
 {M}=\{\rho_\theta | \theta\in\Theta\}.
\end{equation}
Let ${\cal G}_{\rm outlier}$ be a set of possible outlier quantum states.
For example, the set ${\cal G}_{\rm outlier}$ consists of $L$ elements as
\begin{equation}
{\cal G}_{\rm outlier}=\{ \sigma_1,\sigma_2,\ldots,\sigma_L\}.
\end{equation}
The case of a continuous set of outlier quantum states can be defined similarly.

Following the same philosophy as the classical contaminated model \eqref{contam_model},
we define a quantum contaminated model by
\begin{equation}\label{contam_qmodel}
\rho^\varepsilon_\theta= (1-\varepsilon) \rho_\theta+\varepsilon \sigma\quad (\sigma\in{\cal G}_{\rm outlier}).
\end{equation} 
Here, the contamination parameter $\varepsilon\in [0,1)$ represents the strength of contamination.
Importantly, we do not have precise knowledge on the value $\varepsilon$ nor $\sigma$.  
This is in contrast to the conventional noise model in the quantum estimation theory, where we assume a parametric family of noise models. Without knowing a specific form of noisy quantum states, we cannot apply the standard methodology of parametric inference to infer the parameters of interest. 
The simplest case is when there is one particular outlier quantum state $\sigma$, which may or may not be known to us. 
In this situation, we have $\rho^\varepsilon_\theta= (1-\varepsilon) \rho_\theta+\varepsilon \sigma$. 
When the outlier quantum states exhibit a probabilistic structure,
the second term in Eq.~\eqref{contam_qmodel} is expressed as a convex mixture of outlier states. 
Hence, we can express the above model \eqref{contam_qmodel} as
\begin{equation} \label{contam_qmodel2}
\rho^\varepsilon_\theta= (1-\varepsilon) \rho_\theta+\varepsilon \int p(s) \sigma(s) ds,
\end{equation}
where $p(s)$ is a probability density function for the occurrence of the parametric family of outlier states $\{\sigma(s)\}$. 

A few remarks about our setting are as follows.
First, a question of the optimal estimator. 
It is in general a hard task to derive an optimal estimator $\hat{\theta}$ for our model \eqref{contam_qmodel}. 
This is because the MLE, which is asymptotically optimal, is no longer optimal 
in the presence of unknown outlier quantum states. 

Second, an additional complication comes in the quantum case due to the measurement degree of freedom.
In the quantum estimation theory, we can derive an optimal measurement strategy to extract the maximum information
about the parameter $\theta$ from the ideal model ${M}$.
However, this optimal measurement is no longer optimal for the quantum contaminated model \eqref{contam_qmodel}. 
In fact, it is almost impossible to identify the optimal measurement, 
when the set ${\cal G}_{\rm outlier}$ is not fixed but has uncertain elements.
The purpose of robust statistics is to estimate the parameter of interest $\theta$ in the presence of nuisance parameter $\varepsilon$ and unknown outlier quantum states.

With this in mind, we shall not explore an optimal estimation strategy in this paper, but we fix a good measurement for the ideal state.
We then apply the method of the M-estimator to make robust estimates on $\theta$.

Third, the model \eqref{contam_qmodel2} can be understood as a quantum channel (a CP-TP map). 
This quantum noise model acts on the ideal state $\rho_\theta$.
The simplest instance of a single outlier state is expressed as $\rho^\varepsilon_\theta= (1-\varepsilon) \rho_\theta+\varepsilon \sigma$.
Note that this quantum channel is not unital in general.
Although this class of non-unital maps is well studied, the more general forms \eqref{contam_qmodel} or \eqref{contam_qmodel2} are not explored in view of robust statistics to our best knowledge. 

Last, measurement outcomes. A measurement is described by a positive operator-valued measure (POVM).
Let $\Pi_x\ge0$ ($x\in{\cal X}$) be elements of the POVM such that $\int \Pi_x dx=I$ (The identity operator).
When we perform this POVM on the state \eqref{contam_qmodel2},
the resulting measurement outcomes obey the probability density function:
\begin{align} \nonumber
f_\theta^\varepsilon &= \tr\left( \rho^\varepsilon_\theta \Pi_x\right)\\
&=(1-\varepsilon) \tr\left( \rho_\theta\Pi_x\right)+\varepsilon \int p(s)  \tr\left(\sigma(s)\Pi_x\right) ds.
\end{align}
This is a probability mixture of the ideal measurement outcome $\tr\left( \rho_\theta\Pi_x\right)$ and noisy outcomes $\tr\left(\sigma(s)\Pi_x\right)$. This relation establishes connection to the classical contaminated model \eqref{contam_model}.

\subsection{Quantum gaussian state with outliers}
We now consider a concrete example for a quantum contaminated model for a coherent state.
In our model, the ideal state is an unknown coherent state, and outlier quantum states are given by the thermal gaussian states.
A motivation for considering this model is that there occur noisy thermal states with some frequency at the state preparation stage.
This could be treated as the quantum contaminated model of the form \eqref{contam_qmodel}.

The standard definition of the coherent state, which is characterized by a complex number $\alpha\in \mathbb{C}$, is
\begin{equation}\label{def_mu1}
\ket{\alpha}= e^{-\frac{|\alpha|^{2}}{2}} \sum_{j=0}^{\infty} \frac{\alpha^{j}}{\sqrt{j !}}\ket{j},
\end{equation}
with $\ket{j}$ the Fock state of $j$ photons.
We can also define the coherent state by a unitary shift as follows.
Let $\hat{a}^{\dagger}$ and $\hat{a}$ be the creation and annihilation operator,
and define the shift operator by
\begin{equation}
U(\alpha)=e^{\alpha a^{\dagger}-\alpha^{*} a}\quad(\alpha\in \mathbb{C}).
\end{equation}
The coherent state \eqref{def_mu1} is then expressed as a shifted state by $\alpha$ from the vacuum state $\ket{0}$:
\begin{equation}\label{def_mu2}
 \ket{\alpha}=U(\alpha)\ket{0}.
\end{equation}

Next, we consider thermal states as outlier quantum states.
The thermal state at the inverse temperature $\beta=\frac{1}{k_{B} T}$ ($k_{B}$: The Boltzmann constant) is defined by
\begin{equation}\label{def_thermalstate}
\rho^{\rm thermal}_{\beta}=C_{\beta} \sum_{j=0}^{\infty} e^{-\beta j}|j \rangle \langle j|,
\end{equation}
where $C_{\beta}=1-e^{-\beta}$ 
and we set the unit engergy for the single-mode field without loss of generality, i.e., $\hbar \omega=1$.  
Note that the zero temperature limit, $\beta \rightarrow \infty$, converges to the vacuum state:
\begin{equation}
\lim _{\beta \rightarrow \infty} \rho^{\rm thermal}_{\beta}=|0 \rangle\langle0|.
\end{equation}
A quantum gaussian shift state at the inverse temperature $\beta$ is defined by
\begin{equation}\label{def_Qgauss_state}
\rho^{\rm GS}_{\beta, \alpha}=U(\alpha) \rho_{\beta} U(\alpha)^{\dagger},
\end{equation}
which is also expressed in the integral form as
\begin{equation}\label{Qgauss_integral}
\rho^{\rm GS}_{\beta, \alpha}=\frac{1}{\pi} \int  e^{-\frac{|z-\alpha|^{2}}{2 k^{2}}}|z\rangle\langle z| d^{2} z,
\end{equation}
with $2 \kappa^{2}=\left(e^{\beta}-1\right)^{-1}$.
It is straightforward to see the zero temperature limit of the quantum gaussian shift state
is a coherent state, i.e.,
\begin{equation}
\lim _{\beta \rightarrow \infty} \rho^{\rm GS}_{\beta,\alpha}=|\alpha\rangle\langle \alpha|.
\end{equation}
The parameter $\kappa$ ($\kappa\ge0$) represents a dispersion of thermal spreads of coherent states as seen from Eq.~\eqref{Qgauss_integral}.
In the following we mainly use $\kappa$ and denote the quantum gaussian state state simply by $\rho_{\alpha,\kappa}$,
 since there is the one-to-one correspondence between $\beta$ and $\kappa$.

Let $\alpha\in\mathbb{C}$ be the complex parameter of interest, and consider quantum outlier states
described by the quantum gaussian shift state $\rho_{z,\kappa} $.
With these definitions, our quantum contaminated model is defined by
\begin{equation}\label{thermal_outlier_state}
\rho^{\varepsilon}_\alpha = (1-\varepsilon) |\alpha\rangle\langle \alpha| + \varepsilon \int p(z,\kappa) \rho_{z,\kappa} d^2z\,d\kappa .
\end{equation}
In this expression, $p(z,\kappa) $ describes the probability density function of the outlier quantum state $\rho_{z,\kappa}$.
From this expression, we see that the quantum contaminated model is a mixed quantum gaussian models in our setting.
In practice, we take independent density function as $p(z,\kappa)=p_1(z_{\rm R})p_2(z_{\rm I})p_3(\kappa) $,
where $z_{\rm R}=\Re z$ and $z_{\rm I}=\Im z$ denote the real and imaginary part of $z$, respectively.
To implement our numerical simulation of the above model, we consider two different settings as follows.

\noindent
{\bf Single outlier quantum state:}
When there is one particular outlier quantum state, the model \eqref{thermal_outlier_state} is expressed as
\begin{equation} \label{qcontam_single}
\rho^\varepsilon_\alpha = (1-\varepsilon) |\alpha\rangle\langle \alpha| + \varepsilon\rho_{z_0,\kappa_0}.
\end{equation}
This corresponds to $p_{z_0,\kappa_0}(z,\kappa)=\delta(z_R-\Re z_0)\delta(z_I-\Im z_0)\delta(\kappa-\kappa_0)$, where $\delta(x)$ denotes the Dirac delta function.
In this model, our interest is two real parameters, $\theta_{1}=\Re \alpha$ and $\theta_{2}=\Im \alpha$.
All other parameters, $\varepsilon$, $z_0$, and $\kappa_0$, are the nuisance parameters.

\noindent
{\bf Distributed outlier quantum states:}
Consider the case when a center of possible outlier quantum states $\alpha$ are generated by the normal distributions on the phase space
with a given dispersion $\kappa_0$. The quantum contaminated model is expressed as
\begin{equation} \label{qcontam_dist}
\rho^\varepsilon_\alpha = (1-\varepsilon) |\alpha\rangle\langle \alpha|
+ \varepsilon \int f^{\cal N}_{\mu_1,\sigma_1}(z_{\rm R})f^{\cal N}_{\mu_2,\sigma_2}(z_{\rm I}) \rho_{z,\kappa_0} d^2z.
\end{equation}
This corresponds to set a probability distribution for outlier quantum states as $p_{\mu_1,\mu_2,\kappa_0}(z,\kappa)=f^{\cal N}_{\mu_1,\sigma_1}(z_{\rm R})f^{\cal N}_{\mu_2,\sigma_2}(z_{\rm I}) \delta(\kappa-\kappa_0)$ in Eq.~\eqref{thermal_outlier_state}.
To remind ourselves, $f^{\cal N}_{\mu,\sigma}(x)$ is a probability density function of the normal distribution with the expectation value $\mu$ and the standard deviation $\sigma$. Thus, taking a limit $\sigma_1, \sigma_2 \to 0$ reduces to the case of a single outlier quantum state. 
The above distributed outlier quantum states can be easily generalized to the case of multiple centers by adding more terms in Eq.~\eqref{qcontam_dist}.

%We further have two subclasses of this noisy quantum gaussian state model.
%When outlier quantum states are generated a distribution whose expectation value is zero,
%we call them as {\it symmetric outlier quantum states}. The corresponding model is call a {\it symmetric quantum contaminated model}.
%If, on the other hand, outlier states are not symmetrically distributed, we call the model as an {\it asymmetric quantum contaminated model}.
%Within our setting, the choice $p_{0,0,\kappa_0}$ produces a symmetric quantum contaminated model,
%whereas non-vanishing of $\mu_1$ and $\mu_2$ yields an asymmetric quantum contaminated model.

\subsection{Homodyne measurement on the noisy quantum gaussian states}
Denote the standard quadrature operators by
\begin{equation*}
\hat{X}=\frac{1}{\sqrt{2}}\left(\hat{a}+\hat{a}^{\dagger}\right)\mbox{ and }
\hat{P}=\frac{1}{\sqrt{2} i}\left(\hat{a}-\hat{a}^{\dagger}\right).
\end{equation*}
Homodyne measurement at $\phi$ is defined by a projection measurement of an observable,
\begin{equation}
\hat{X}_{\phi}=\hat{X} \cos \phi+ \hat{P} \sin \phi.
\end{equation}
As is well known in quantum optics, a homodyne measurement at $\phi$ on the coherent state $\ket{\alpha}$
gives statistics of the normal distribution whose probability density function is
\begin{equation}
\operatorname{P}_{\alpha}(x | \phi)=\sqrt{\frac{2}{\pi}} 
\exp\left[-2\left(x-\Re\left(\alpha e^{-i \phi}\right)\right)^{2}\right].
\end{equation}
Therefore, denoting $\theta_{1}=\Re \alpha$ and $\theta_{2}=\Im \alpha$,
respectively, it is equivalent to the normal distribution:
\begin{equation*}
\label{gauss_homodyn}
{\cal N}\left(\theta_{1} \cos \phi+\theta_{2} \sin \phi, \frac{1}{2}\right),
\end{equation*}
where ${\cal N}(\mu,\sigma)$ denotes the normal distribution with the expectation value $\mu$ and the standard deviation $\sigma$.
More generally, statistics of measurement outcomes of homodyne measurement on the quantum gaussian state $\rho_{\alpha,\kappa}$ \eqref{def_Qgauss_state} is
\begin{equation} \label{stat_homodyne}
\operatorname{P}_{\alpha,\kappa}(x | \phi)=\frac{1}{\sqrt{2 \pi\left(\kappa^{2}+1\right)}} 
\exp\left[-\frac{1}{2\left(\kappa^{2}+\frac{1}{4}\right)}\left(x- \Re\left(\alpha e^{-i \phi}\right)\right)^{2}\right] .
\end{equation}
This identifies statistics of the normal distribution:
\begin{equation*}
{\cal N}\left(\theta_{1} \cos \phi+\theta_{2} \sin \phi, \sqrt{\kappa^{2}+\frac{1}{4}}\right).
\end{equation*}

Combining our quantum contaminated model \eqref{thermal_outlier_state} and measurement statistics \eqref{stat_homodyne},
we obtain statistics for homodyne measurement at $\phi$ on the noisy quantum gaussian states as
%\begin{multline}
\begin{equation}
  \label{contaminated_model_dist}
\operatorname{P}_{\alpha,\varepsilon}(x | \phi)
=(1-\varepsilon) \operatorname{P}_{\alpha}(x | \phi)
+\varepsilon \int p_{z_0,\kappa_0}(z,\kappa)  \operatorname{P}_{z,\kappa}(x | \phi)  d^2z\,d\kappa .
\end{equation}
%\end{multline}
This is a convex mixture of normal distributions.
For the case of a single outlier quantum state \eqref{qcontam_single}, we have
\begin{equation}
\operatorname{P}_{\alpha,\varepsilon}(x | \phi)
=(1-\varepsilon) \operatorname{P}_{\alpha}(x | \phi)+\varepsilon \operatorname{P}_{z_0,\kappa_0}(x | \phi) .
\end{equation}
When we consider homodyne measurement on a more general quantum contaminated model \eqref{qcontam_dist},
measurement statistics is given by
\begin{align}
\operatorname{P}_{\alpha,\varepsilon}(x | \phi)
=(1-\varepsilon) \operatorname{P}_{\alpha}(x | \phi)+\varepsilon \int f^{\cal N}_{\mu_1,\sigma_1}(z_{\rm R})f^{\cal N}_{\mu_2,\sigma_2}(z_{\rm I})
\operatorname{P}_{z,\kappa_0}(x | \phi) d^2z.
%\operatorname{P_{\alpha,\varepsilon}}(x | \phi)
%&=(1-\varepsilon) \operatorname{P}_{\alpha}(x | \phi)\\
%&\hfill +\varepsilon \int f^{\cal N}_{\mu_1,\sigma_1}(z_{\rm R})f^{\cal N}_{\mu_2,\sigma_2}(z_{\rm I})
%\operatorname{P}_{z,\kappa_0}(x | \phi) d^2z.
%&=(1-\varepsilon) \operatorname{P}_{\alpha}(x | \phi)\\
%&\hfill +\varepsilon \int f^{\cal N}_{\mu_1,\sigma_1}(z_{\rm R})f^{\cal N}_{\mu_2,\sigma_2}(z_{\rm I})
%\operatorname{P}_{z,\kappa_0}(x | \phi) d^2z.
\end{align}
%To derive the last line, we used the convolution property of the normal distribution,
%\begin{equation}
%\int f^{\cal N}_{\mu_1,\sigma}(z) f^{\cal N}_{\mu_1,\sigma}(x) dx=
%\end{equation}
We note that the second term can be further simplified by the use of a convolution formula of the normal distribution.
For our purpose, the above expression suffices to implement our numerical simulation,
which will be discussed in the next section.

\section{Phase estimation of noisy coherent state}\label{sec5}
\begin{figure}[H]
\begin{center}
\includegraphics[width=0.55\linewidth]{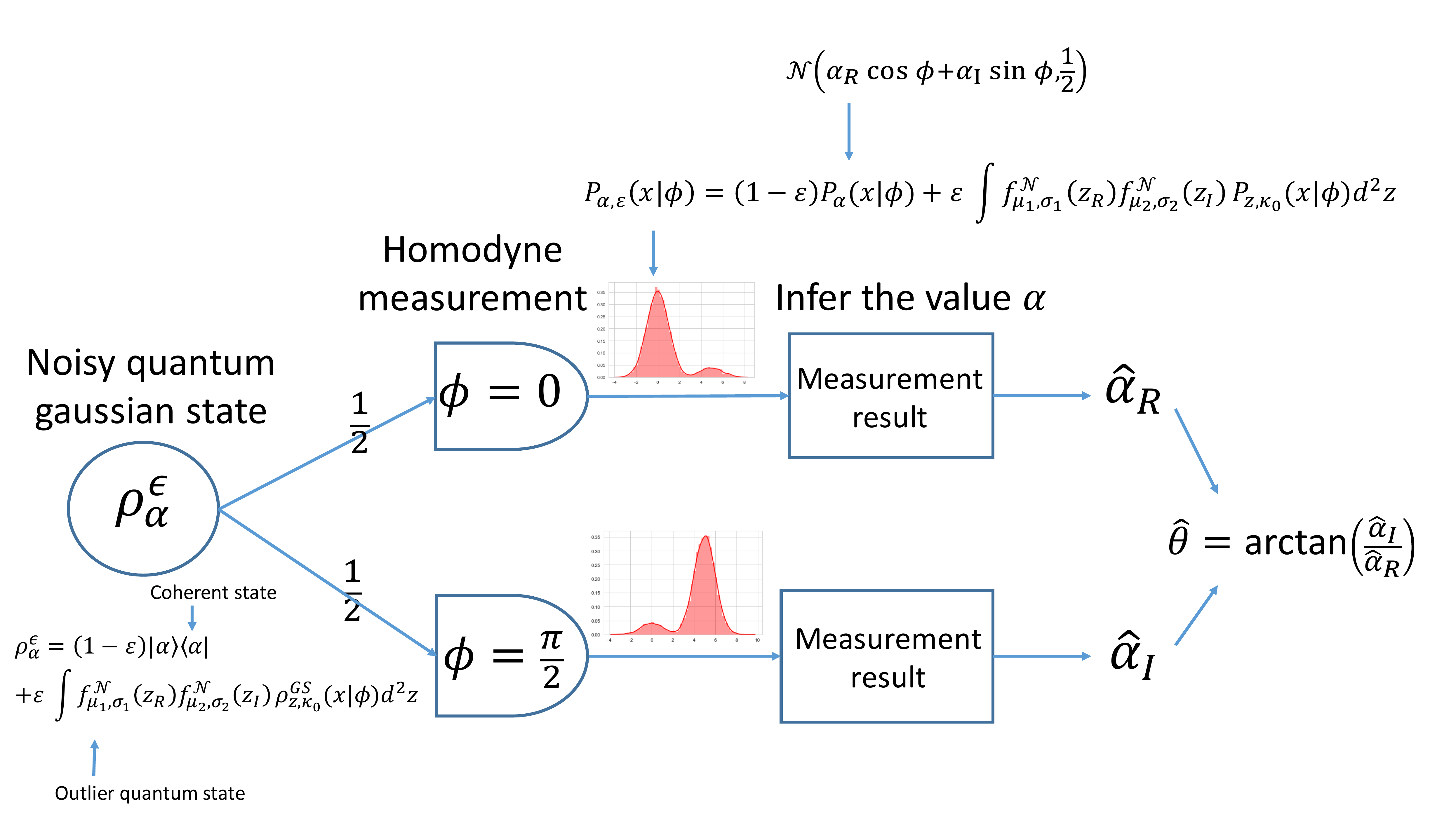}
\caption{A schematic diagram of our model and estimation procedure.}
\label{fig_setting}
\end{center}
\end{figure}

To illustrate advantages of the M-estimator in noisy quantum gaussian systems,
we consider a phase estimation problem, which has many important applications for quantum information processing protocols \cite{aspachs2009phase,pinel2013quantum,bradshaw2018ultimate,oh2019optimal,lee2019using,Arnhem_2019,oh2020optimal,assad2020accessible}.
Suppose an unknown coherent state $\ket{\alpha}$ is given.
Importantly, the ideal state is parametrized by two real parameters $(\alpha_{R},\alpha_{I})\in\mathbb{R}^2$.
($\alpha_{R}=\Re\alpha$ and $\alpha_{I}=\Im\alpha$)
Our primary interest to estimate the phase $\theta=\arctan \left(\frac{\alpha_{I}}{\alpha_{R}}\right)$
of an unknown coherent state $\ket{\alpha}$.
In our setting, the phase $\theta$ is the parameter of interest and the other parameter, the amplitude of the state,
$r=\sqrt{\alpha_{R}^{2}+\alpha_{I}^{2}}$ is the nuisance parameter of the model \cite{suzuki2020quantum,suzuki2020nuisance}.
The optimal estimation strategy for this problem is known.
However, this optimal measurement depends on the unknown phase $\phi$ and cannot be implemented.
In the following, we consider a random mixture of two homodyne measurements at $\phi=0,\pi/2$.
As an estimator, we first apply M-estimators $(\hat{\alpha}_{I},\hat{\alpha}_{R})$ to infer the value $\alpha$.
We then convert it to phase by
\begin{equation} \label{estimator_theta}
\hat{\theta}=\arctan \left(\frac{\hat{\alpha}_{I}}{\hat{\alpha}_{R}}\right).
\end{equation}
A schematic diagram of our setting is given in Figure~\ref{fig_setting}. 

\subsection{Numerical simulation}
We describe procedures of our numerical simulation.
We set the true coherent state as $\alpha=10+4i$.
The true phase value is $\theta=\arctan(0.4)\simeq 0.3805$.
We randomly generate $n$ quantum gaussian states, which is a convex mixture of the ideal coherent state and outlier quantum states.
(Either single outlier quantum state case \eqref{qcontam_single} or distributed outlier quantum state case \eqref{qcontam_dist}.)
We perform a random homodyne measurements at $\phi=0,\pi/2$ with equal probability.
From measurement outcomes $\mathbf{x}$ at $\phi=0$, we apply an M-estimator to estimate $ \hat{\alpha}_{R}(\mathbf{x})$.
The imaginary part will be estimated by homodyne measurement at $\phi=\pi/2$.
We repeat the iterative algorithm of Section~\ref{sec:IterAlg} to obtain estimates. 
We stop iteration when the difference between two successive estimates is below a predetermined threshold value. 
We then apply the formula \eqref{estimator_theta} to obtain an estimate $\hat {\theta} _ {n}(\mathbf{x})$. 
In our simulation, we change the sample size $n$ for a given contamination parameter $\varepsilon$. 
We compare two types M-estimators (bisquare and gamma) together with the sample mean and the median. 
(The sample mean for the ideal coherent state corresponds to the MLE in our setting.) 
%A schematic diagram of our setting is given in Figure~\ref{fig_setting}.

\subsection{Single outlier quantum state}
\begin{figure}[htbp]
  \begin{minipage}[b]{0.475\linewidth}
    \centering
    \includegraphics[width=\linewidth]{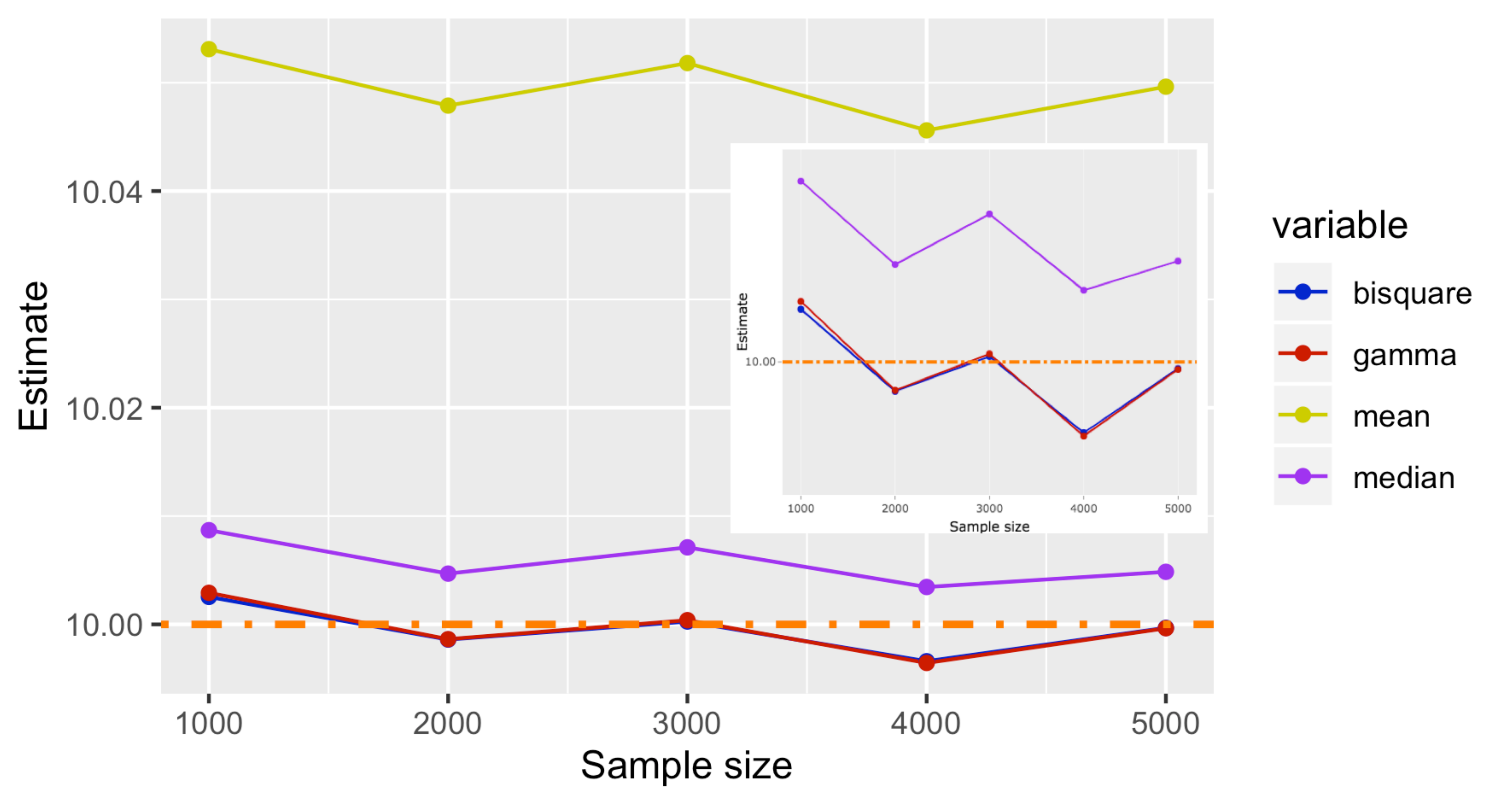}
    \subcaption{Comparison of estimators for $\alpha_{R}$.}\label{fig2a}
  \end{minipage}
  \begin{minipage}[b]{0.475\linewidth}
    \centering
    \includegraphics[width=\linewidth]{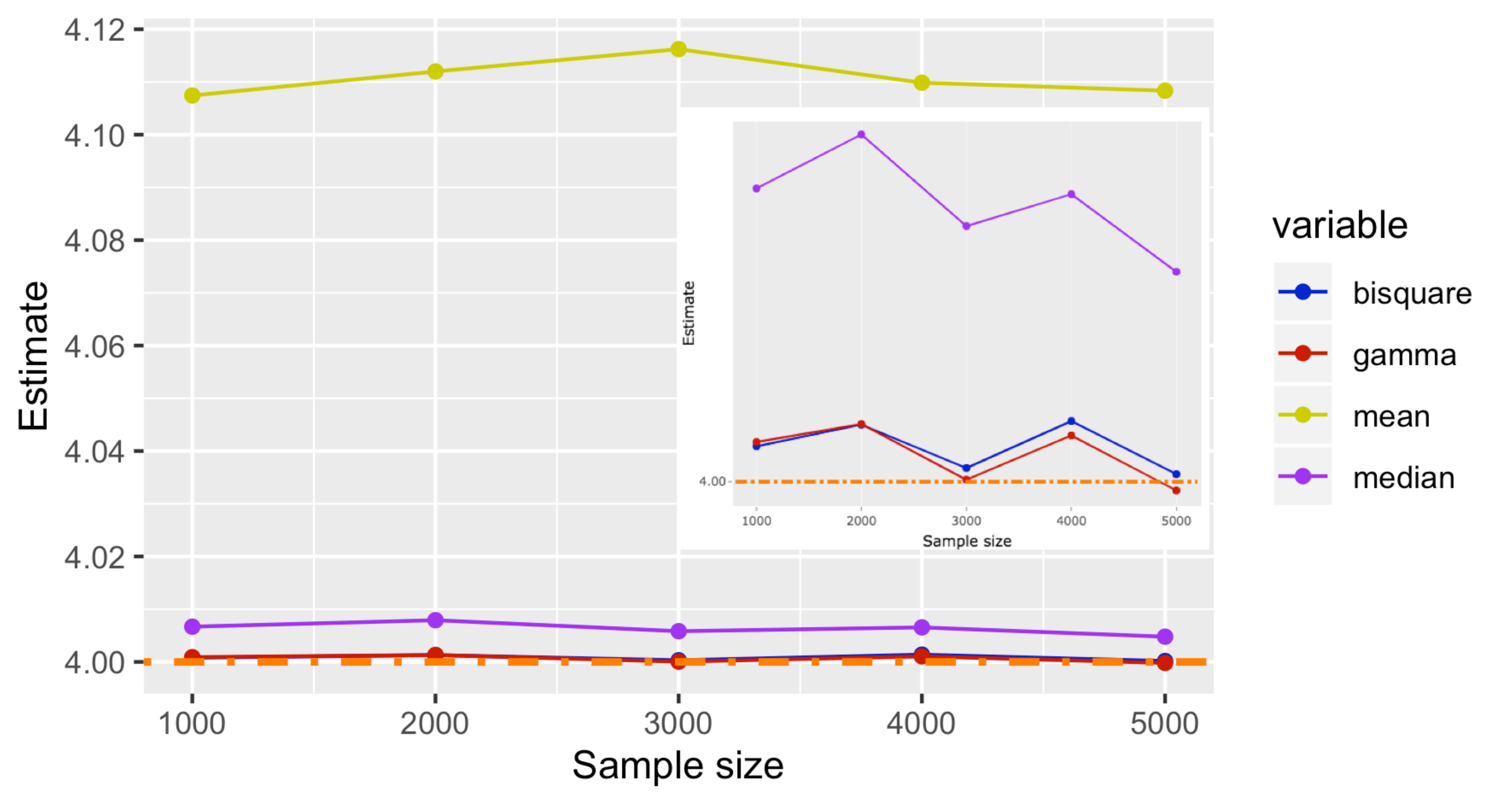}
    \subcaption{Comparison of estimators for $\alpha_{I}$.}\label{fig2b}
  \end{minipage}\\ %\\
  \begin{minipage}[b]{0.475\linewidth}
    \centering
    \includegraphics[width=\linewidth]{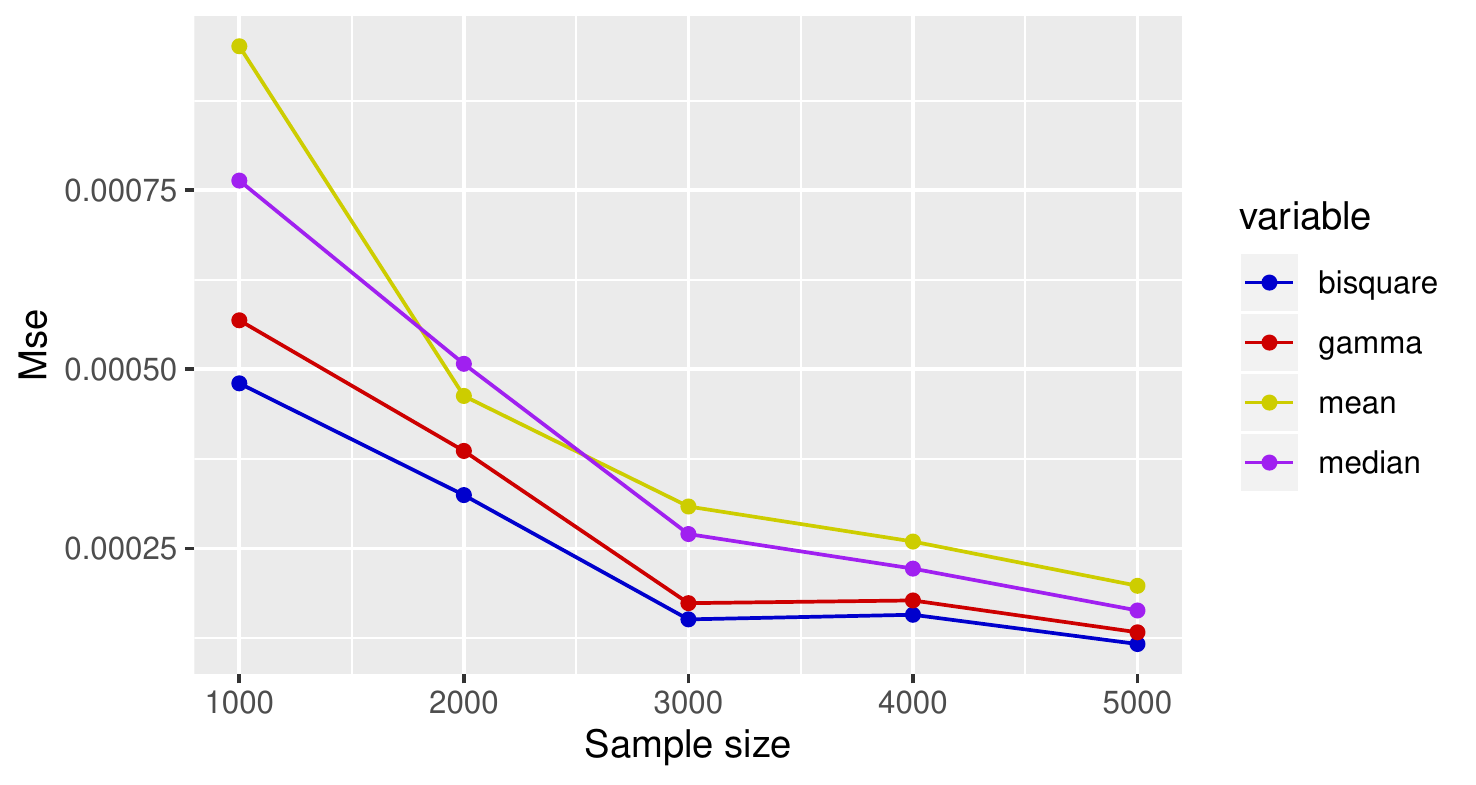}
    \subcaption{Comparison of MSEs for $\alpha_{R}$.}\label{fig2c}
  \end{minipage}
  \begin{minipage}[b]{0.475\linewidth}
    \centering
    \includegraphics[width=\linewidth]{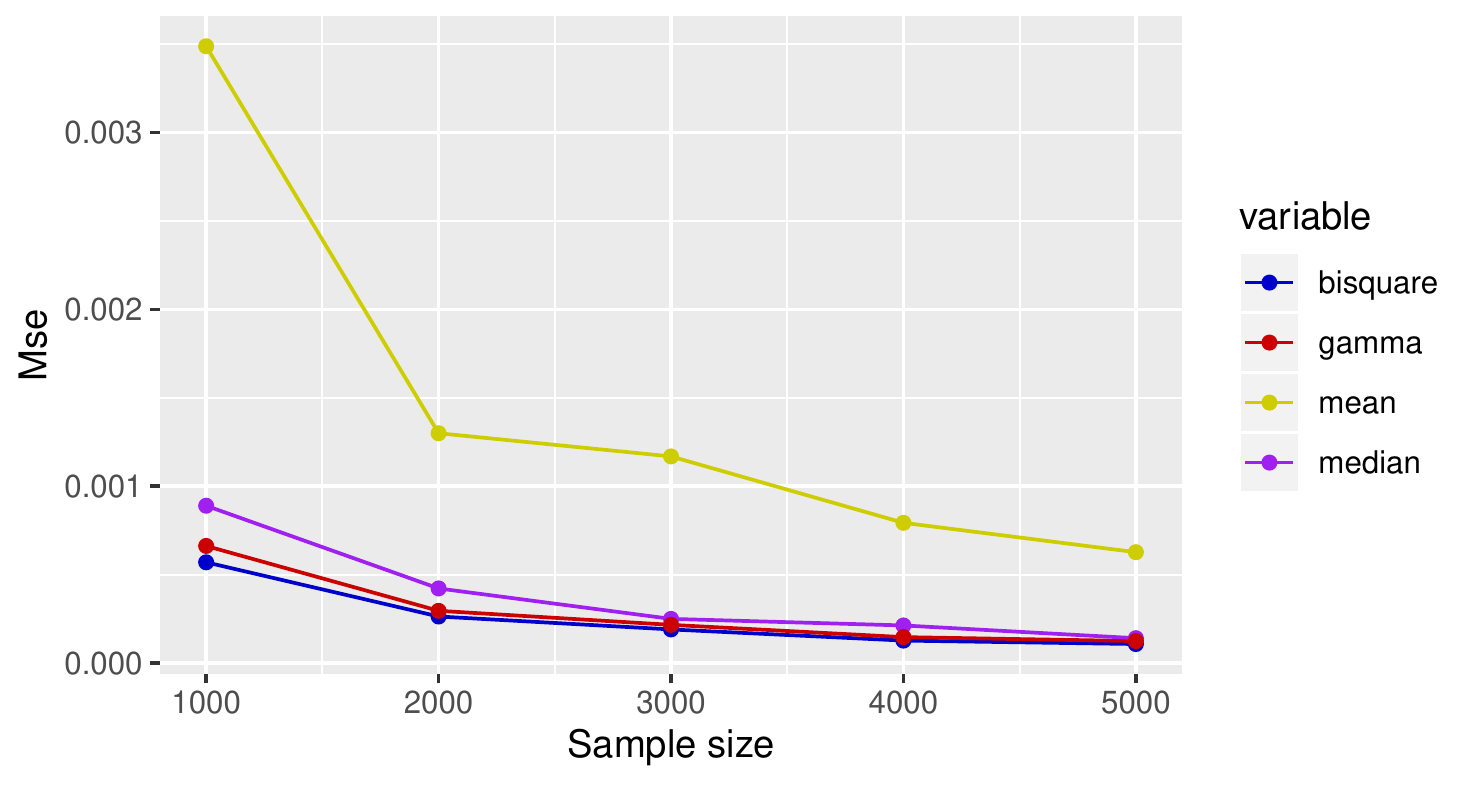}
    \subcaption{Comparison of MSEs for $\alpha_{I}$.}\label{fig2d}
  \end{minipage}
  \caption{Performances of M-estimators (bisquare and gamma) and the standard estimators (mean and median) as functions of the sample size.}\label{fig2}
\end{figure}

An outlier quantum state is set as $z_{0}=15+15i,\kappa_{0}=0.1$ in Eq.~\eqref{qcontam_single}.
We first change the sample size $n=1000, 2000,\ldots,5000$ to analyze performances of M-estimators for the contamination parameter $\varepsilon=0.01$.
Comparison of M-estimators $ \hat{\alpha}_{R}$ and $ \hat{\alpha}_{I}$ vs the sample size $n$ is plotted in Figures~\ref{fig2a} and \ref{fig2b}, respectively.
In Figures~\ref{fig2c} and \ref{fig2d}, we plot the mean square errors (MSEs) of these estimators vs the sample size $n$ for the contamination parameter $\varepsilon=0.01$.
In these figures, plotted are estimates from the sample mean (yellow-green), the median (purple), bisquare M-estimator (blue), and gamma M-estimator (red), which are averaged over 500 runs of each setting. The true values are plotted by dashed-dotted line (orange).
From Figures~\ref{fig2a} and \ref{fig2b}, we see that bisqure and gamma M-estimators both performs well when compared to the sample mean and the median. 
The observed differences between these two figures come from the fact that ${\alpha}_{R}$ is harder to estimate than ${\alpha}_{I}$. This is because measurement outcomes at $\phi=0$ are more sensitive to outlier quantum states as two gaussian distributions are close to each other. 
Good performances in the MSEs are also observed for bisqure and gamma M-estimators in Figures~\ref{fig2c} and \ref{fig2d}.
We should stress that the mean and the median are not consistent estimators for our setting.

\begin{figure}[htbp]
  \begin{minipage}[b]{0.475\linewidth}
    \centering
    \includegraphics[width=\linewidth]{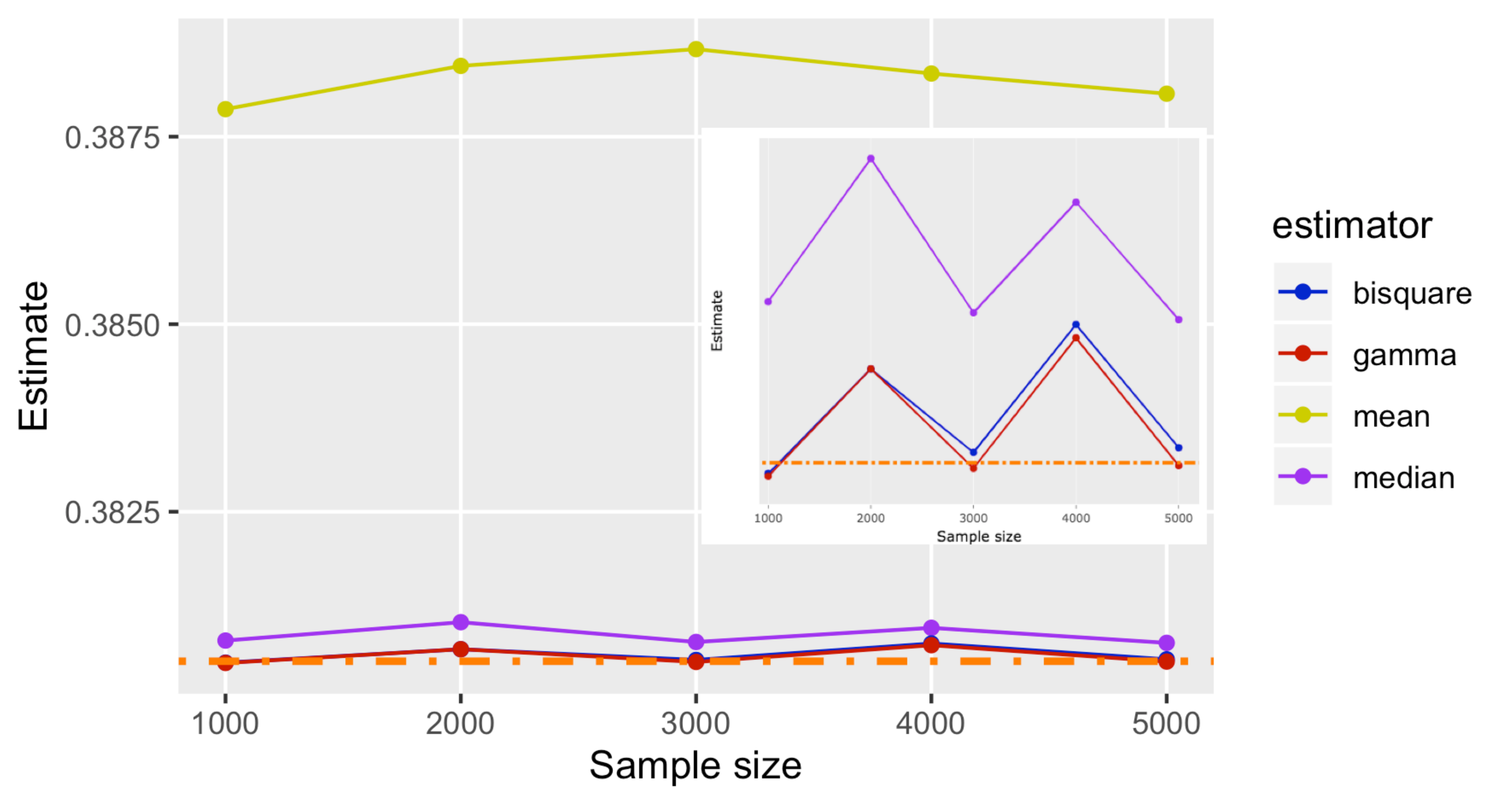}
    \subcaption{Comparison of estimators for $\theta$.}\label{fig3a}
  \end{minipage}
  \begin{minipage}[b]{0.475\linewidth}
    \centering
    \includegraphics[width=\linewidth]{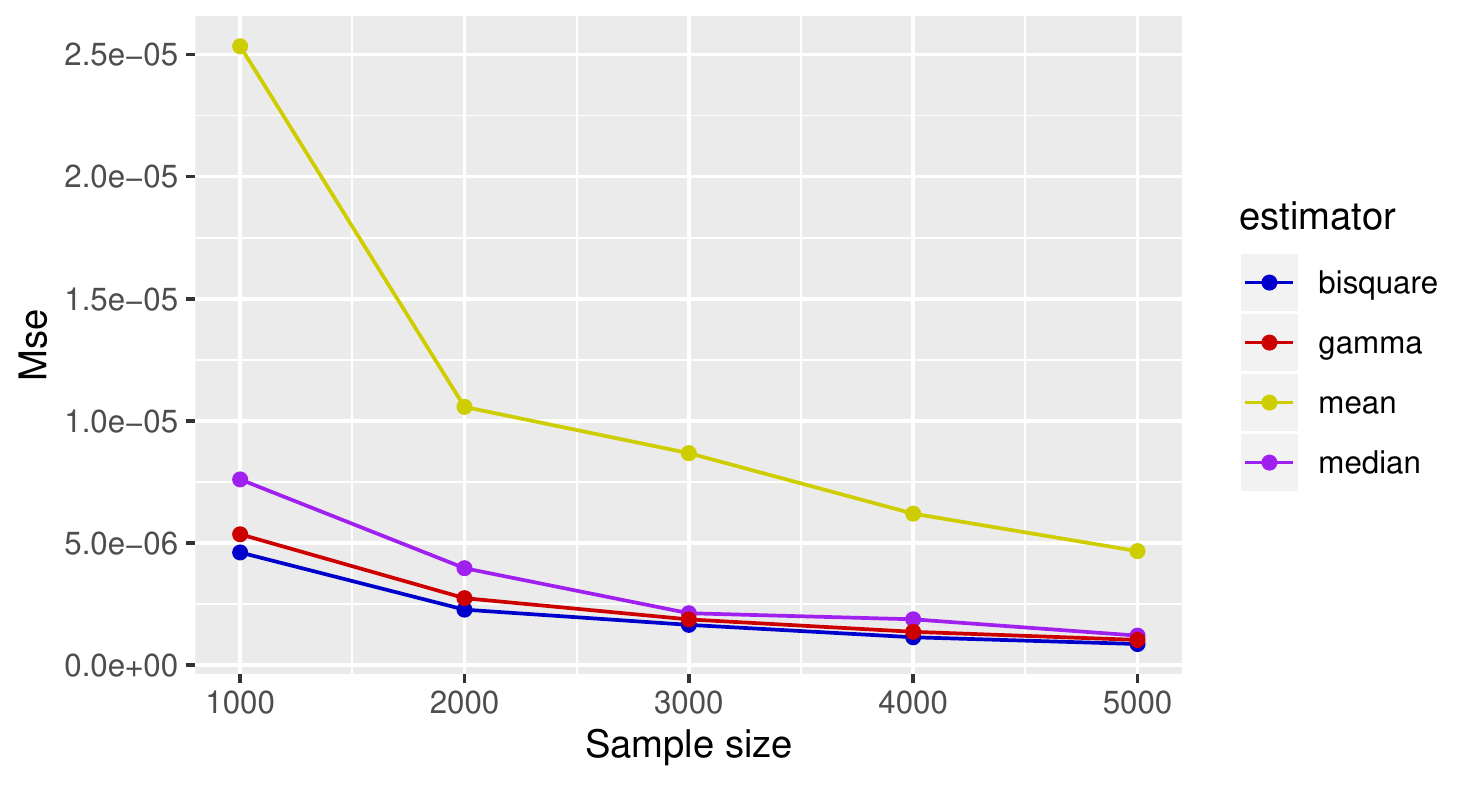}
    \subcaption{Comparison of MSE for $\theta$.}\label{fig3b}
  \end{minipage}
  \caption{Performances of M-estimators for phase of a coherent state in the presence of a single outlier quantum state.}\label{fig3}
\end{figure}

Next, we plot the result for phase estimation in Figure~\ref{fig3}.
The same convention as in Figure~\ref{fig2} is used. 
From Figures~\ref{fig3a} and \ref{fig3b}, we draw the same conclusion for bisqure and gamma M-estimators as in Figure~\ref{fig2}. Namely, they are robust and consistent estimators. 
This is well understood from the fact that it is sufficient to have good estimators for $\alpha_{R}$ and $ \alpha_{I}$ to estimate phase accurately. 

\subsection{Robustness of M-estimators}\label{sec:num_robust}
\begin{figure}[htbp]
\centering
\includegraphics[width=0.65\linewidth]{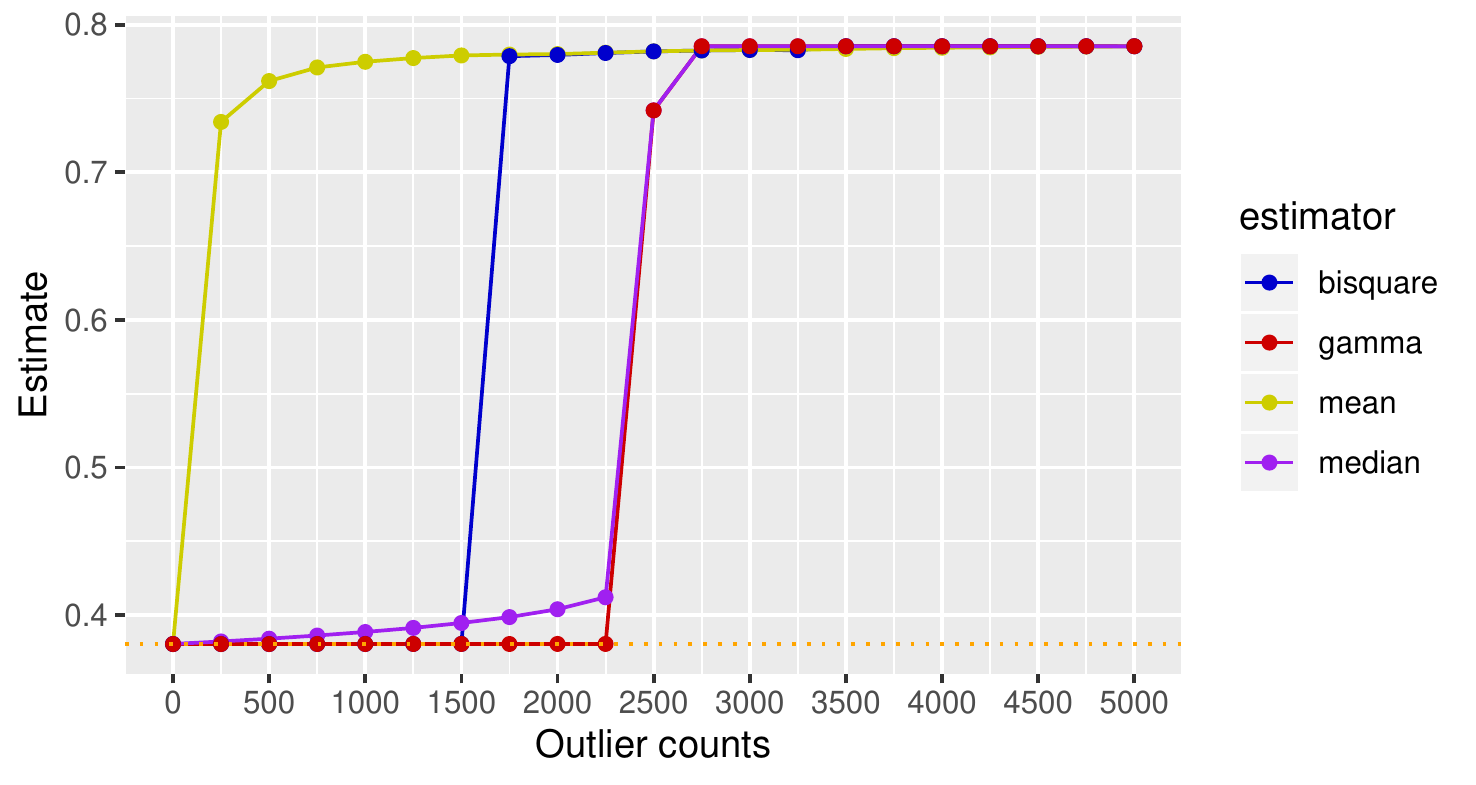}
%\vspace{-0.5cm}
\caption{The finite breakdown point (FBP) for estimating phase of a coherent state.}
\label{fig4}
\end{figure}

We now analyze robustness of two M-estimators, the sample mean, and the median.
First, we analyze the FBP \eqref{def_FBP} for phase estimation in Figure~\ref{fig4}.
To produce this figure, we randomly replace data by artificial data, which are generated by
the normal distribution ${\cal N}(1000,0.1)$.
The step size for increasing the number of replacing outliers (Outlier counts) is $250$ in Figure~\ref{fig4}. 
In this setting, an estimator is regarded as being outside the parameter space when it returns the value $\arctan (1)\simeq 0.785$. This corresponds to $ \hat{\alpha}_{R}$ and $ \hat{\alpha}_{I}$ being close to $1000$. 

Based on this figure, FBPs for four estimators are roughly estimated as follows: mean = 1500/5000 = 0.3, bisquare = 1750/5000 = 0.35, gamma and median = 2750/5000 = 0.55. 
From a first glance at this figure, we might conclude that the median and gamma M-estimator are robust estimators. 
However, we already know from Figure~\ref{fig3} that the median does not show consistency as an estimator.
This discrepancy is easily resolved by realizing that the definition of the FBP is not related to the actual estimate at all.
From this simple example, we conclude that the FBP should not be used as a measure of robustness a priori.

\begin{figure}[H]
\centering
\includegraphics[width=0.65\linewidth]{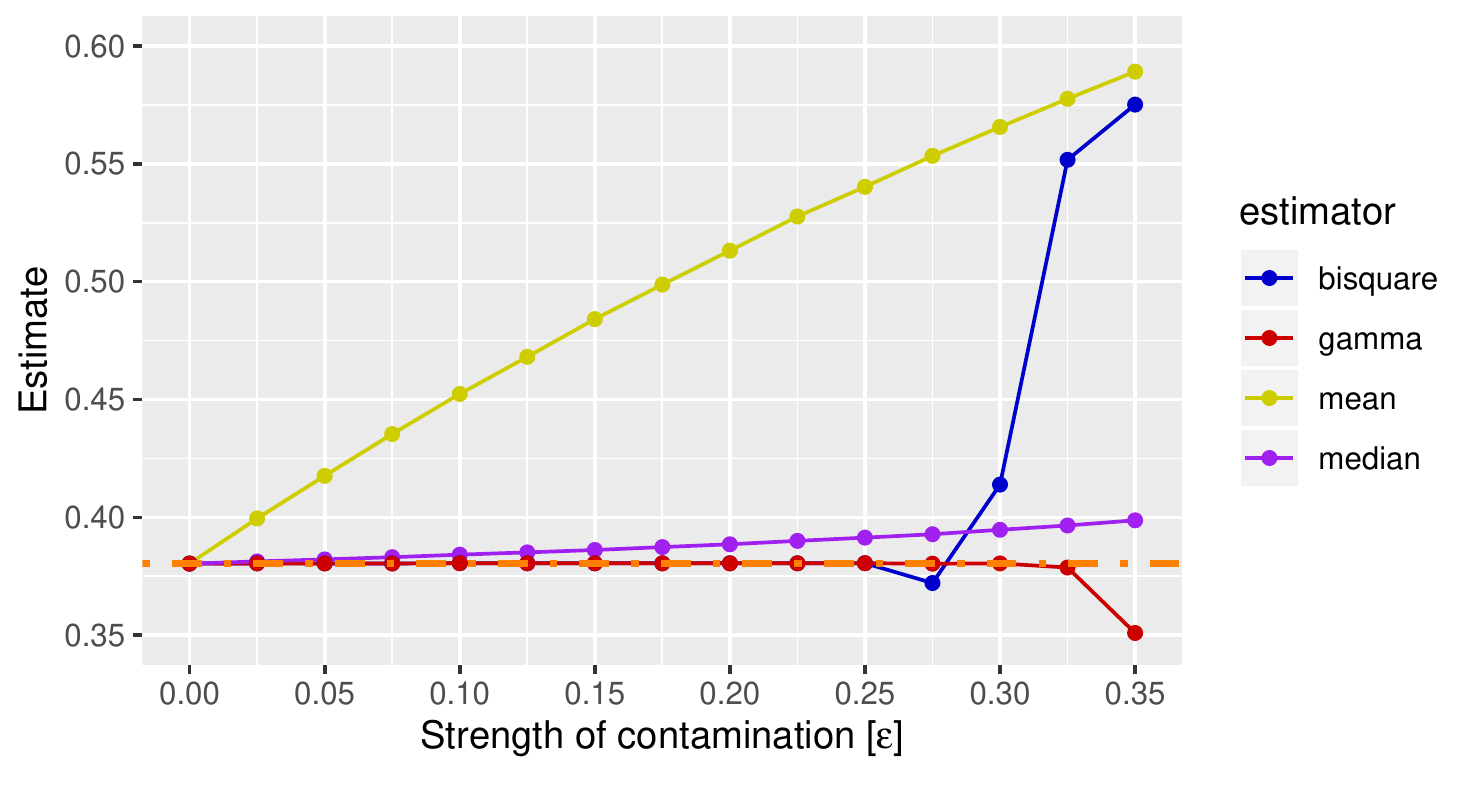}
%\vspace{-0.5cm}
\caption{$\varepsilon$-curve for phase estimation for the sample size $n=5000$.}
\label{fig5}
\end{figure}

Next, we plot the $\varepsilon$-curve proposed at the end of Section~\ref{sec:rubustness}.
This is to plot behaviors of an M-estimator as a function of the contamination parameter $\varepsilon$ for a fixed sample size.
In Figure~\ref{fig5}, we show the $\varepsilon$-curve for phase estimation for the sample size $n=5000$.
From this figure, we find that gamma M-estimators are the most robust estimators for small $\varepsilon$. 
In contrast, the sample mean and the median smoothly deviate from the true value.
However, it is noted that bisquare and gamma M-estimators behave uncontrollable manners after some threshold: $\varepsilon\simeq0.25$ for bisquare and $\varepsilon\simeq0.35$ for gamma.

Before we move to the next subsection, we report the average number of iterations of the M-estimator algorithm in Table~\ref{table1} and Table~\ref{table2}. 
From two tables, we see that the number of iterations to get a robust estimate is relatively small. 
This shows that the M-estimator can be implemented efficiently. 

\begin{table}[H]
    \centering
        \caption{The average number of iterations of robust estimator algorithm for $\alpha_{R}$}
    \scalebox{0.85}{
    \begin{tabular}{|c|l|l|l|l|l|l|l|l|l|l|l|l|l|l|l|}
    \hline
        $\varepsilon$ & 0 & 0.025 & 0.05 & 0.075 & 0.1 & 0.125 & 0.15 & 0.175 & 0.2 & 0.225 & 0.25 & 0.275 & 0.3 & 0.325 & 0.35 \\ \hline
        bisquare & 3.21 & 5.00 & 5.01 & 5.00 & 5.00 & 5.11 & 5.99 & 6.00 & 6.08 & 7.11 & 9.07 & 20.02 & 11.00 & 6.67 & 5.29 \\ \hline
        gamma & 7.97 & 7.99 & 7.99 & 8.00 & 8.02 & 8.04 & 8.12 & 8.28 & 8.57 & 8.85 & 9.00 & 9.20 & 9.89 & 11.60 & 17.77 \\ \hline
%        bisquare & 3.214 & 5.004 & 5.006 & 5 & 5 & 5.114 & 5.988 & 6 & 6.08 & 7.11 & 9.068 & 20.02 & 11.004 & 6.668 & 5.294 \\ \hline
%        gamma & 7.97 & 7.988 & 7.994 & 8.002 & 8.018 & 8.036 & 8.118 & 8.278 & 8.574 & 8.852 & 8.998 & 9.2 & 9.894 & 11.596 & 17.768 \\ \hline
    \end{tabular}
    }
    \label{table1}
\end{table}
\begin{table}[H]
    \centering
    \caption{The average number of iterations of robust estimator algorithm for $\alpha_{I}$}    
    \scalebox{0.85}{
    \begin{tabular}{|c|l|l|l|l|l|l|l|l|l|l|l|l|l|l|l|}
    \hline
        $\varepsilon$ & 0 & 0.025 & 0.05 & 0.075 & 0.1 & 0.125 & 0.15 & 0.175 & 0.2 & 0.225 & 0.25 & 0.275 & 0.3 & 0.325 & 0.35 \\ \hline
        bisquare & 3.19 & 5.00 & 5.00 & 4.96 & 4.05 & 4.67 & 5.00 & 5.00 & 5.41 & 6.06 & 7.07 & 9.64 & 20.06 & 9.99 & 6.50 \\ \hline
        gamma & 7.97 & 7.99 & 8.01 & 8.01 & 8.02 & 8.02 & 8.05 & 8.09 & 8.09 & 8.12 & 8.18 & 8.20 & 8.21 & 8.23 & 8.38 \\ \hline
%        bisquare & 3.194 & 5 & 5 & 4.956 & 4.048 & 4.67 & 5 & 5 & 5.408 & 6.064 & 7.068 & 9.636 & 20.058 & 9.994 & 6.498 \\ \hline
%        gamma & 7.97 & 7.99 & 8.006 & 8.006 & 8.016 & 8.022 & 8.054 & 8.09 & 8.094 & 8.12 & 8.176 & 8.204 & 8.21 & 8.234 & 8.38 \\ \hline
    \end{tabular}
    }
    \label{table2}
\end{table}

\subsection{Distributed outlier quantum states}
\begin{figure}[ht]
\begin{subfigure}{0.475\textwidth}
  \centering
    \includegraphics[width=\columnwidth]{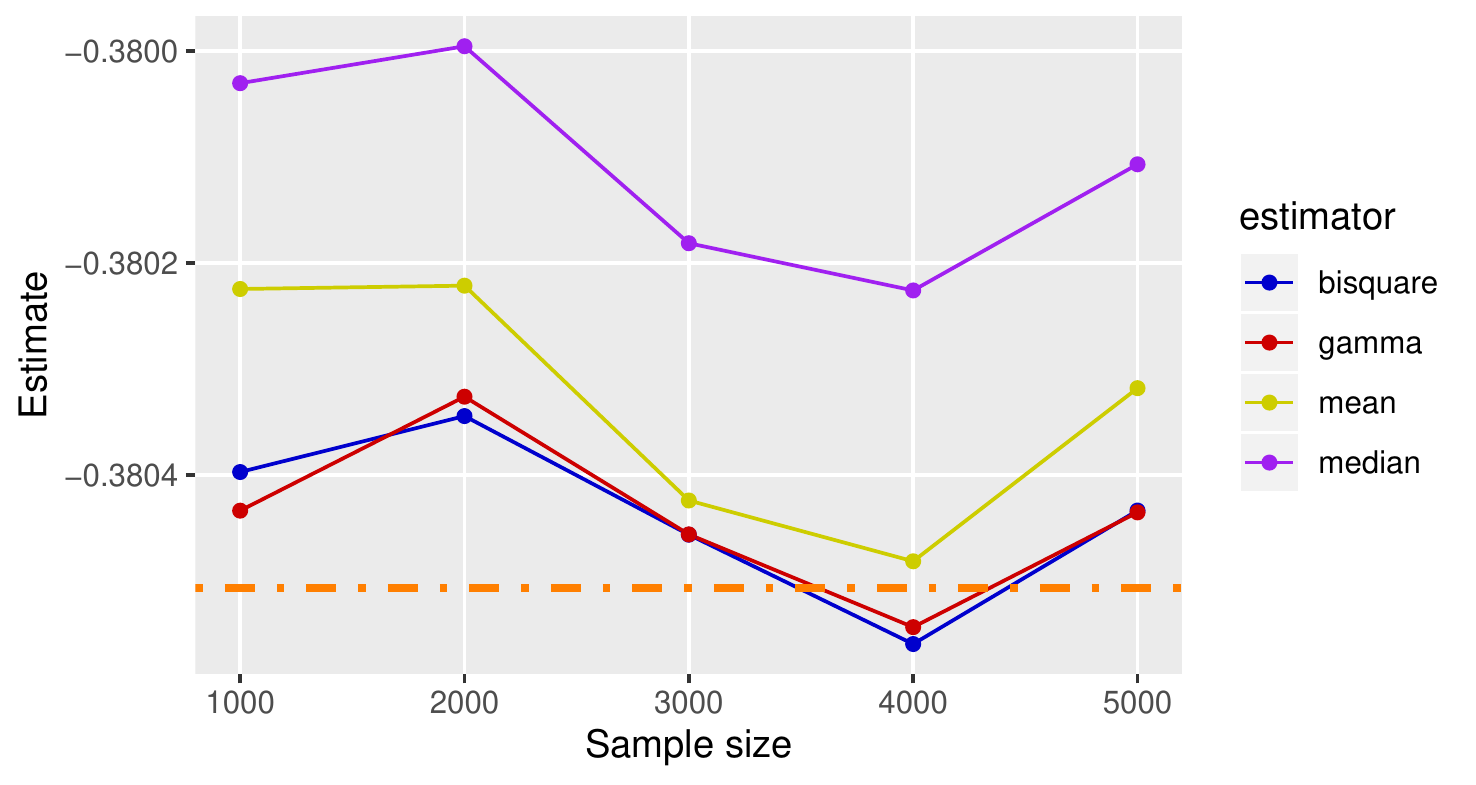}
  \caption{Comparison of estimators for $\theta$.}
  \label{fig6a}
\end{subfigure}
\begin{subfigure}{0.475\textwidth}
  \centering
  \includegraphics[width=\columnwidth]{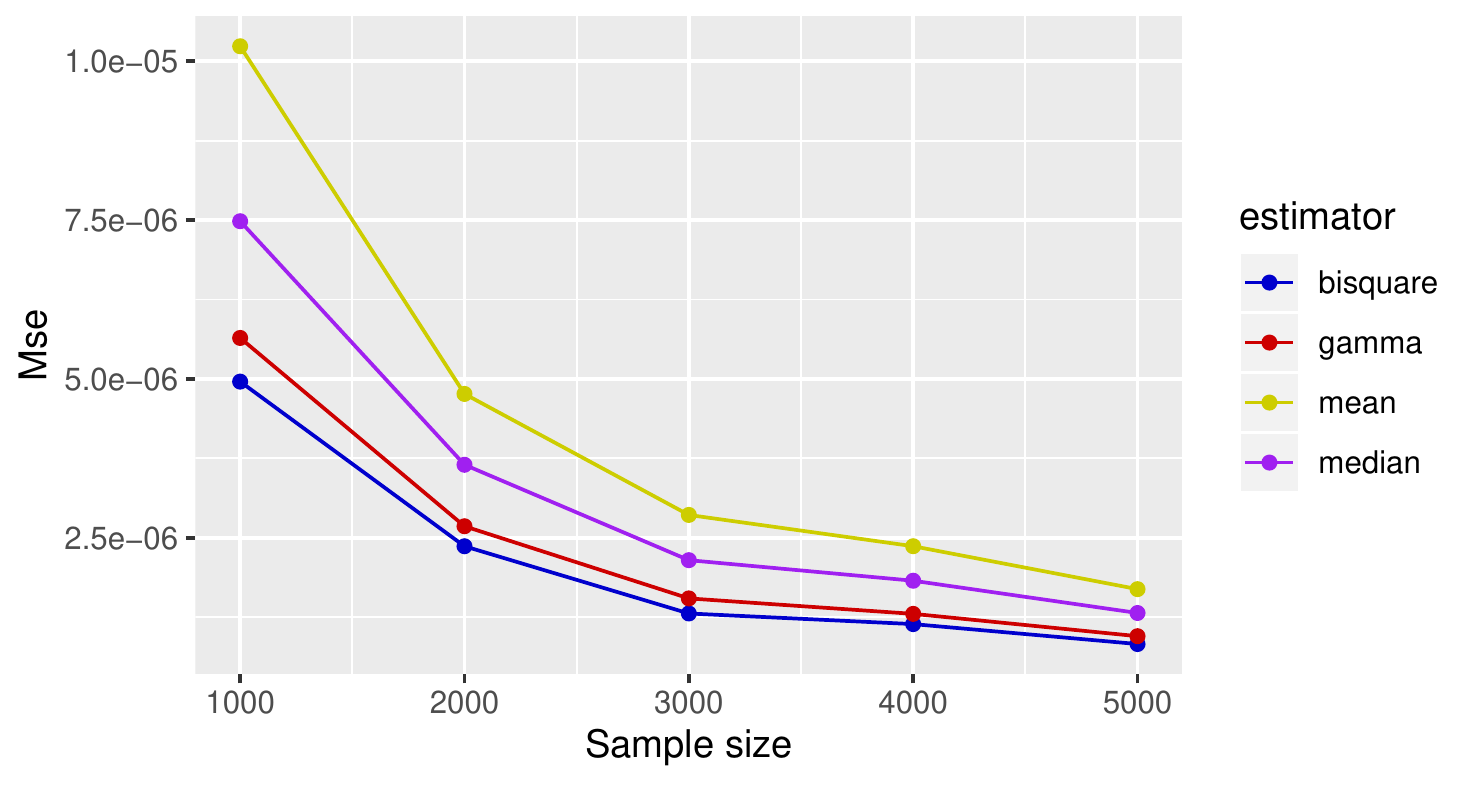}
  \caption{Comparison of MSE for $\theta$.}
  \label{fig6b}
\end{subfigure}
\caption{}
\label{fig6}
\end{figure}

We now consider more general setting where outlier quantum states are distributed around the vacuum state of the form Eq.~\eqref{qcontam_dist}.
Outlier quantum states are generated according to the normal distribution as $\alpha_{R}\sim N(0.1,0.1)$, $\alpha_{I}\sim N(0.1,0.1)$ with a fixed value of dispersion $\kappa_0=0.1$.
We set the true coherent state as $\alpha=10-4 i$ in this example. 
The true phase value is $\theta=\arctan (-0.4) \simeq -0.3805$. 
In Figure~\ref{fig6}, we show performances of M-estimators together with the sample mean and the median for the contamination parameter $\varepsilon=0.01$.
The same convention as in Figure~\ref{fig2} is used.
From Figure~\ref{fig6}, we observe that bisquare and gamma M-estimator perform well compared to other estimators.
In this class of outlier quantum states, the median seems to be a bad choice to use.
In fact, its performance is worse than that of the sample mean.

\begin{figure}[htbp]
\centering
\includegraphics[width=0.65\linewidth]{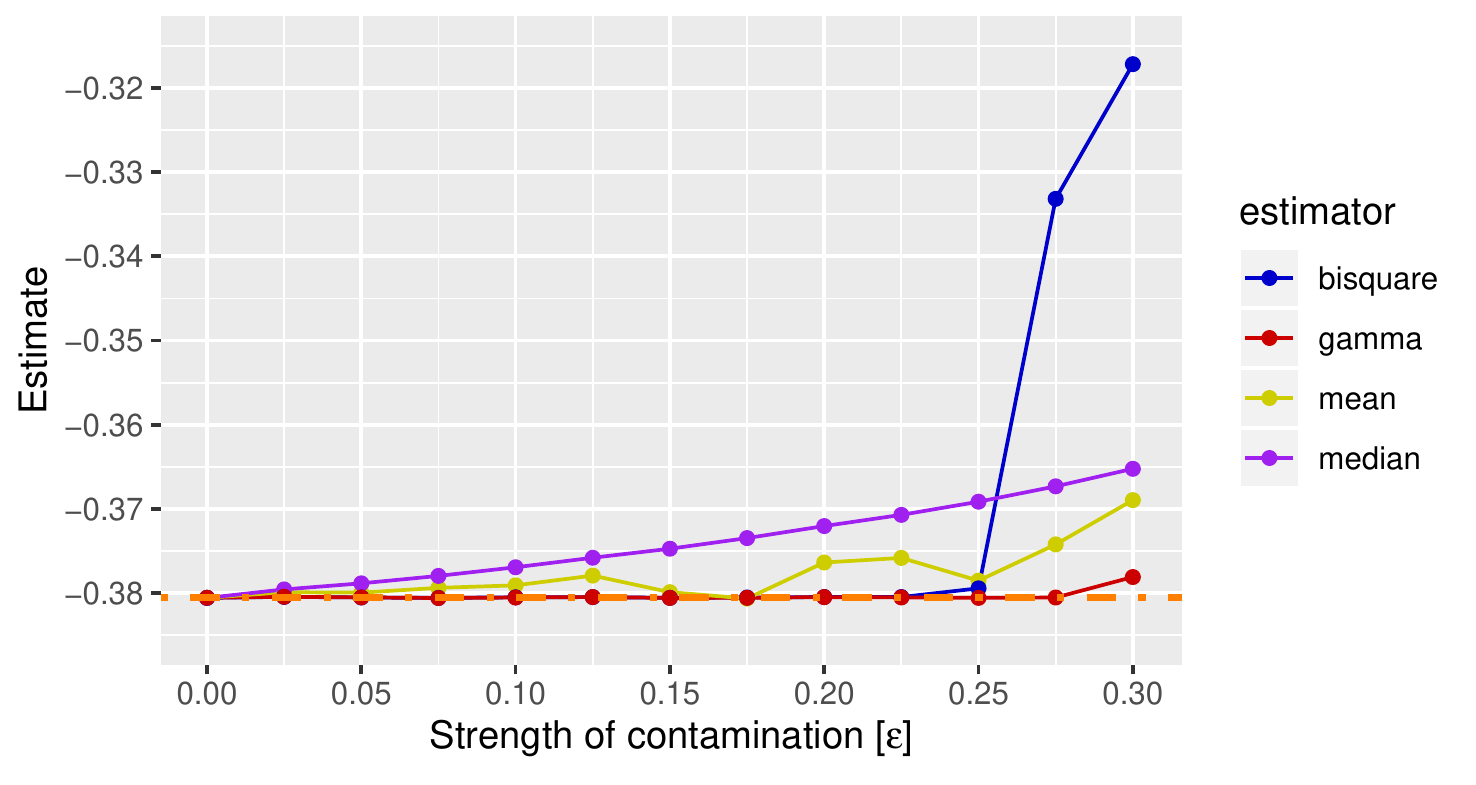}
%\vspace{-0.5cm}
\caption{$\varepsilon$-curve for phase estimation for the sample size $n=5000$.}
\label{fig7}
\end{figure}

Finally, we analyze robustness of M-estimators by the $\varepsilon$-curve.
In Figure~\ref{fig7}, we plot the $\varepsilon$-curve for phase estimation of the sample size $n=5000$.
From this result, we find that gamma M-estimator is most robust and bisquare M-estimator is second.
The sample mean is also found to be relatively robust with small fluctuation. 
However, this behavior in fact depends on the nature of outlier quantum states as discussed in the next subsection. 
Lastly, the result of the median shows that it is not trustable for the distributed outlier quantum states of this kind. 

\subsection{Discussion}
Following two reasons described below, we conclude that the gamma M-estimator is the most robust estimator than the mean, median, and the well-known robust estimator, bisquare, upon estimating the phase $\theta$ of the coherent state. 
First, from Figure~\ref{fig3a} and Figure~\ref{fig6a}, gamma is better than the other estimators in terms of accuracy in estimating the phase $\theta$.
Second, $\varepsilon$-curves (Figure~\ref{fig5} and Figure~\ref{fig7}) suggest that the gamma M-estimator is the best in terms of robustness as well. 

From the MSE plots (Figure~\ref{fig3b} and Figure~\ref{fig6b}), it was confirmed that the gamma estimator and the bisquare estimator have similar convergence speeds.
Furthermore, from Table.~\ref{table1} and Table.~\ref{table2}, it was found that the iteration of the robust estimation algorithm was suppressed to 10 or less for both gamma and bisquare before break down.
In Figure~\ref{fig6a}, relatively good accuracy of the mean estimation is observed. This is because of the effect of the nuisance parameter in our model. In fact, 
the sample means for $\alpha_{R} $and $\alpha_{I}$ behave very poorly with large biases. 
However, the errors of each estimation of $\alpha_{R} $and $\alpha_{I}$ are canceled out. 
This then yields an improvement in estimating the phase $\theta$. 
We remark that this kind of an accidental improvement can only happen in a special case. 
When we do not have prior knowledge on possible outlier quantum states, M-estimators give more robust and reliable estimates. 

\section{Conclusion and outlook}\label{sec6}
In this paper, we have studied the problem of phase estimation for coherent states in the presence of unavoidable outlier quantum states.
Measurement outcomes were then distributed according to a convex mixture of the ideal distribution and noisy distributions due to outliers. 
To overcome this problem, the methodology of M-estimators was introduced based on well-established theory of robust statistics.
Two specific M-estimators, bisquare and gamma M-estimators, were studied to illustrate the advantage over conventional estimators. 
We have found that gamma M-estimator is most accurate as well as robust against the effect of outlier quantum states. 

Before closing our paper, we make a few remarks regarding future works.
Our formalism applies to more general estimation setting with a quantum gaussian state such as squeezed states and entangled states. 
This is done by replacing the ideal state, a coherent state, by other state of interest in Eq.~\eqref{thermal_outlier_state}.
More general quantum contaminated models can be treated without any modification.
Preliminary studies show that M-estimators work efficiently in this case as well.
Our quantum statistical model in the presence of outlier quantum states can also be
used to model imperfection in other elements of quantum information processing such as gate operations, measurement processes, and so on.
In this sense, the proposed method of M-estimators has wider applications to overcome a certain types of SPAM errors in practice.
Last, we mainly consider a quantum communication scenario in this paper.
Utilization of robust statistics should be relevant to handle the problem of outlier quantum states
when implementing other high precision measurement schemes based on quantum resources such as quantum metrology, quantum imaging, and quantum sensing.

\section*{Acknowledgment}
The work is partly supported by JSPS KAKENHI grant number JP17K05571.


\begin{thebibliography}{37}

\bibitem{andrews2015robust}
D.~Andrews and F.~Hampel, {\em Robust Estimates of Location: Survey and Advances}, 
Princeton Legacy Library, Princeton University Press, 2015.

\bibitem{Arnhem_2019}
M.~Arnhem, E.~Karpov, and N.~J. Cerf, 
{\it Optimal estimation of parameters encoded in quantum coherent state quadratures}, 
{Applied Sciences}, {\bf 9}(20), 4264, 2019.

\bibitem{aspachs2009phase}
M.~Aspachs, J.~Calsamiglia, R.~Mu{\~n}oz-Tapia, and E.~Bagan, 
{\it Phase estimation for thermal gaussian states}, 
{Physical Review A}, {\bf 79}(3), 033834, 2009.

\bibitem{assad2020accessible}
S.~M. Assad, J.~Li, Y.~Liu, N.~Zhao, W.~Zhao, P.~K. Lam, Z.~Ou, and X.~Li, 
{\it Accessible precisions for estimating two conjugate parameters using gaussian probes}, 
{Physical Review Research}, {\bf 2}(2), 023182, 2020.

\bibitem{bradshaw2018ultimate}
M.~Bradshaw, P.~K. Lam, and S.~M. Assad, 
{\it Ultimate precision of joint quadrature parameter estimation with a gaussian probe}, 
{Physical Review A}, {\bf 97}(1), 012106, 2018.

\bibitem{braunstein2005quantum}
S.~L. Braunstein and P.~Van~Loock.
{\it Quantum information with continuous variables}, 
{Reviews of Modern Physics}, {\bf 77}(2), 513, 2005.

\bibitem{camerer2018evaluating}
C.~F. Camerer, A.~Dreber, F.~Holzmeister, T.-H. Ho, J.~Huber, M.~Johannesson,
  M.~Kirchler, G.~Nave, B.~A. Nosek, T.~Pfeiffer, et~al.
{\it Evaluating the replicability of social science experiments in nature and science between 2010 and 2015}, 
{Nature Human Behaviour}, {\bf 2}(9), 637--644, 2018.

\bibitem{d2000parameter}
G.~M. D'ariano, M.~G. Paris, and M.~F. Sacchi, 
{\it Parameter estimation in quantum optics}, 
{Physical Review A}, {\bf 62}(2), 023815, 2000.

\bibitem{donoho1983notion}
D.~L. Donoho and P.~J. Huber, 
{\it The notion of breakdown point}, 
{A festschrift for Erich L. Lehmann}, 157184, 1983.

\bibitem{ferrie2014self}
C.~Ferrie, 
{\it Self-guided quantum tomography}, 
{Physical review letters}, {\bf 113}(19), 190404, 2014.

\bibitem{fujisawa2008robust}
H.~Fujisawa and S.~Eguchi, 
{\it Robust parameter estimation with a small bias against heavy contamination}, 
{Journal of Multivariate Analysis}, {\bf 99}(9), 2053--2081, 2008.

\bibitem{fujiwara1999estimation}
A.~Fujiwara and H.~Nagaoka, 
{\it An estimation theoretical characterization of coherent states}, 
{Journal of Mathematical Physics}, {\bf 40}(9), 4227--4239, 1999.

\bibitem{hampel2011robust}
F.~R. Hampel, E.~M. Ronchetti, P.~J. Rousseeuw, and W.~A. Stahel, 
{\em Robust statistics: the approach based on influence functions}, volume 196, 
John Wiley \& Sons, 2011.

\bibitem{hayashi2017quantum}
M.~Hayashi, 
{\em Quantum information theory}, 
{Graduate Texts in Physics, Springer}, 2017.

\bibitem{helstrom1968minimum}
C.~W. Helstrom, 
{\it The minimum variance of estimates in quantum signal detection}, 
{IEEE Transactions on information theory}, {\bf 14}(2), 234--242, 1968.

\bibitem{helstrom1969quantum}
C.~W. Helstrom, 
{\it Quantum detection and estimation theory}, 
{Journal of Statistical Physics}, {\bf 1}(2), 231--252, 1969.

\bibitem{helstrom}
C.~W. Helstrom, 
{\em Quantum detection and estimation theory}, Academic press, 1976.

\bibitem{helstrom1974noncommuting}
C.~W. Helstrom and R.~Kennedy, 
{\it Noncommuting observables in quantum detection and estimation theory}, 
{IEEE Transactions on Information Theory}, {\bf 20}(1), 16--24, 1974.

\bibitem{holevo}
A.~S. Holevo, 
{\em Probabilistic and statistical aspects of quantum theory}, Edizioni della Normale, 2011.

\bibitem{huber1984finite}
P.~J. Huber, 
{\it Finite sample breakdown of m-and p-estimators}, 
{The Annals of Statistics}, 119--126, 1984.

\bibitem{huber2004robust}
P.~J. Huber, {\em Robust statistics}, volume 523, John Wiley \& Sons, 2004.

\bibitem{lee2019using}
C.~Lee, C.~Oh, H.~Jeong, C.~Rockstuhl, and S.-Y. Lee, 
{\it Using states with a large photon number variance to increase quantum fisher information in single-mode phase estimation}, 
{Journal of Physics Communications}, {\bf 3}(11), 115008, 2019.

\bibitem{maronna2019robust}
R.~A. Maronna, R.~D. Martin, V.~J. Yohai, and M.~Salibi{\'a}n-Barrera, 
{\em Robust statistics: theory and methods (with R)}, John Wiley \& Sons, 2019.

\bibitem{merkel2013self}
S.~T. Merkel, J.~M. Gambetta, J.~A. Smolin, S.~Poletto, A.~D. C{\'o}rcoles,
  B.~R. Johnson, C.~A. Ryan, and M.~Steffen, 
{\it Self-consistent quantum process tomography}, 
{Physical Review A}, {\bf 87}(6), 062119, 2013.

\bibitem{oh2020optimal}
C.~Oh, C.~Lee, S.~H. Lie, and H.~Jeong, 
{\it Optimal distributed quantum sensing using gaussian states}, 
{Physical Review Research}, {\bf 2}(2), 023030, 2020.

\bibitem{oh2019optimal}
C.~Oh, C.~Lee, C.~Rockstuhl, H.~Jeong, J.~Kim, H.~Nha, and S.-Y. Lee, 
{\it Optimal gaussian measurements for phase estimation in single-mode gaussian metrology}, 
{npj Quantum Information}, {\bf 5}(1), 1--9, 2019.

\bibitem{QSEbook}
M.~G.~A. Paris and J.~E. \v{R}eh\'a\v{c}ek, {\em Quantum State Estimation}, Springer, 2004.

\bibitem{pinel2013quantum}
O.~Pinel, P.~Jian, N.~Treps, C.~Fabre, and D.~Braun, 
{\it Quantum parameter estimation using general single-mode gaussian states}, 
{Physical Review A}, {\bf 88}(4), 040102, 2013.

\bibitem{serafini2017quantum}
A.~Serafini, 
{\em Quantum continuous variables: a primer of theoretical methods}, CRC press, 2017.

\bibitem{sugiyama2018reliable}
T.~Sugiyama, S.~Imori, and F.~Tanaka, 
{\it Reliable characterization of super-accurate quantum operations} arXiv Preprint:1806.02696, 2018.

\bibitem{suzuki2020nuisance}
J.~Suzuki, 
{\it Nuisance parameter problem in quantum estimation theory: Tradeoff relation and qubit examples}, 
{Journal of Physics A: Mathematical and Theoretical}, {\bf 53}(26), 264001, 2020.

\bibitem{suzuki2020quantum}
J.~Suzuki, Y.~Yang, and M.~Hayashi, 
{\it Quantum state estimation with nuisance parameters}, 
{Journal of Physics A: Mathematical and Theoretical}, 2020. (forthcoming)

\bibitem{wang2007quantum}
X.-B. Wang, T.~Hiroshima, A.~Tomita, and M.~Hayashi, 
{\it Quantum information with gaussian states}, 
{Physics reports}, {\bf 448}(1-4), 1--111, 2007.

\bibitem{wasserstein2016pvalue}
R.~L. Wasserstein and N.~A. Lazar, 
{\it The asa statement on p-values: Context, process, and purpose}, 
{The American Statistician}, {\bf 70}(2), 129--133, 2016.

\bibitem{wasserstein2019beyond}
R.~L. Wasserstein, A.~L. Schirm, and N.~A. Lazar, 
{\it Moving to a world beyond ``$p<0.05$"}, 
{The American Statistician}, {\bf 73}(sup1), 1--19, 2019.

\bibitem{wilcox2011introduction}
R.~R. Wilcox, 
{\em Introduction to robust estimation and hypothesis testing}, Academic press, 2011.

\bibitem{yuen1973multiple}
H.~Yuen and M.~Lax, 
{\it Multiple-parameter quantum estimation and measurement of nonselfadjoint observables}, 
{IEEE Transactions on Information Theory}, {\bf 19}(6), 740--750, 1973.

\end{thebibliography}
\end{document}